\documentclass[lettersize,journal]{IEEEtran}
\usepackage{amsmath,amsfonts}
\usepackage{algorithm}
\usepackage{array}
\usepackage[caption=false,font=normalsize,labelfont=sf,textfont=sf]{subfig}
\usepackage{textcomp}
\usepackage{stfloats}
\usepackage{url}
\usepackage{verbatim}
\usepackage{cite}
\usepackage{subfig}
\usepackage[table,xcdraw]{xcolor}
\usepackage{float} 
\usepackage{amsmath,amssymb,amsfonts}
\usepackage{algorithmic}
\usepackage{multirow}
\usepackage{graphicx}
\usepackage{textcomp}
\usepackage{xcolor}
\usepackage[table,xcdraw]{xcolor}
\usepackage{tabto}
\usepackage{threeparttable}
\usepackage{algorithm}
\usepackage{tablefootnote}
\usepackage{breqn}
\usepackage{lipsum}
\usepackage{multirow}
\usepackage{colortbl}
\usepackage{hhline}
\def\BibTeX{{\rm B\kern-.05em{\sc i\kern-.025em b}\kern-.08em
    T\kern-.1667em\lower.7ex\hbox{E}\kern-.125emX}}

\def\x{{\mathbf x}}

\def\x{{\bf x}}

\begin{document}

\title{Bias-Scalable Near-Memory CMOS Analog Processor for Machine Learning}
\author{\IEEEauthorblockN{Pratik Kumar$^\dagger$, Ankita Nandi$^\dagger$, Shantanu Chakrabartty$^{\ast}$, Chetan Singh Thakur$^\dagger$}\thanks{$^\star $This work is now accepted in IEEE for publication with DOI: 10.1109/JETCAS.2023.3234570. Copyright may be transferred without notice, after which this version may no longer be accessible.}
\\
\IEEEauthorblockA{\{pratikkumar, ankitanandi, csthakur\}@iisc.ac.in, \{shantanu\}@wustl.edu \\
$^\dagger$Department of Electronic Systems Engineering, Indian Institute of Science, Bangalore, India, 560012\\
$^{\ast}$Department of Electrical and Systems Engineering, Washington University in St. Louis,USA, 63130}

}



\maketitle

\begin{abstract}
Bias-scalable analog computing is attractive for implementing machine learning (ML) processors with distinct power-performance specifications. For instance, ML implementations for server workloads are focused on higher computational throughput for faster training, whereas ML implementations for edge devices are focused on energy-efficient inference. In this paper, we demonstrate the implementation of bias-scalable approximate analog computing circuits using the generalization of the margin-propagation principle called shape-based analog computing (S-AC). The resulting S-AC core integrates several near-memory compute elements, which include: (a) non-linear activation functions; (b) inner-product compute circuits; and (c) a mixed-signal compressive memory, all of which can be scaled for performance or power while preserving its functionality. Using measured results from prototypes fabricated in a 180nm CMOS process, we demonstrate that the performance of computing modules remains robust to transistor biasing and variations in temperature. In this paper, we also demonstrate the effect of bias-scalability and computational accuracy on a simple ML regression task. 
\end{abstract}


\section{Introduction}

\begin{figure}[t]
	\centering
		\subfloat[]{\includegraphics[width=0.42\linewidth]{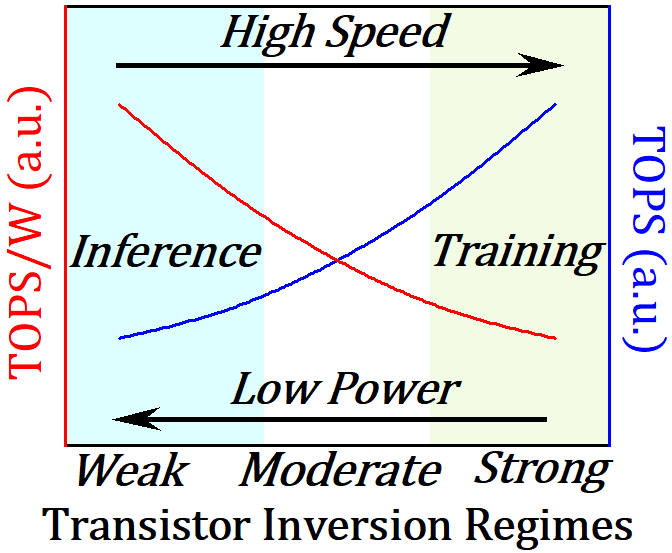}
		\label{tops}}
	\hfil
	\subfloat[]{\includegraphics[width=0.55\linewidth]{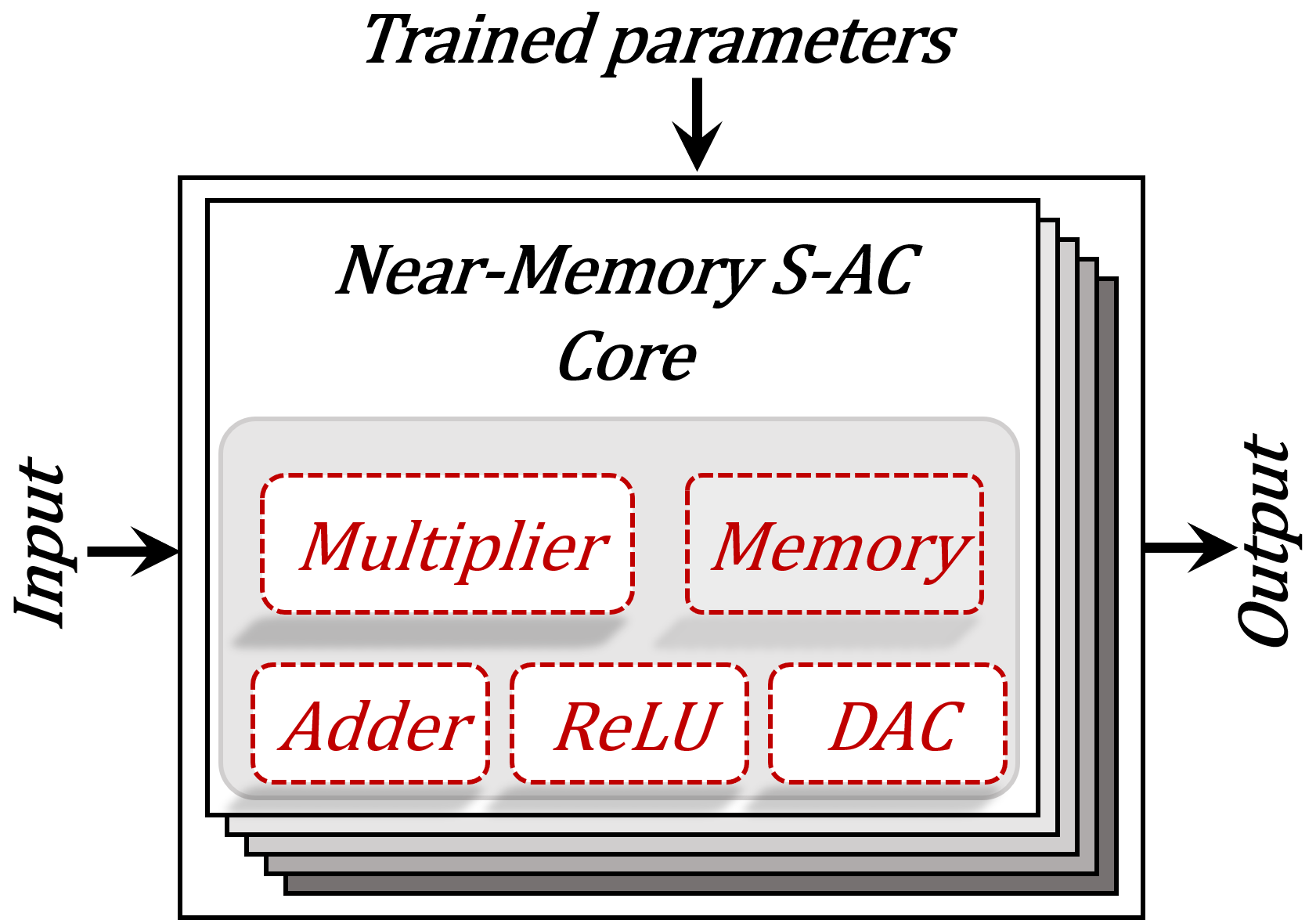}
		\label{blockML_b}}
	\vfil
\caption{\protect\subref{tops}~Desired performance efficiency ($TOPS/W$) and  operational speed plot for a system that can be tuned to execute both edge application and server workload; \protect\subref{blockML_b}~S-AC based near-memory ML core depicting basic computational unit such as Multiplier, DAC, ReLU, Adder along with local memory elements (for storing trained parameters) required for efficient ML computation.}
\end{figure}

\IEEEPARstart{A}{nalog} computing offers a novel paradigm for designing machine learning (ML) systems~\cite{forbes_ibm,demler2018mythic,chakrabarttyGert,thakur2016low,gupta2019low} because the circuits can exploit computational primitives inherent in the device physics along with conservation principles to achieve very high computational density and energy efficiency. However, conventional analog computing circuits operate within a pre-defined transistor biasing regime (weak-inversion~\cite{translinear_weak} or strong-inversion~\cite{translinear_strong}) to ensure sufficient dynamic range and compliance with respect to temperature variations. This approach limits scaling the design across applications that demand distinct power-performance specifications. An illustration of this trade-off is highlighted in Fig.~\ref{tops}. It can be observed from Fig.~\ref{tops} that when the transistors are biased in strong-inversion (SI), higher speed can be achieved but at the cost of increased power consumption, signifying a lower TOPS/W (Trillions Operations per Second per Watt). In weak-inversion (WI), higher energy efficiency (or higher TOPS/W) can be achieved but at the expense of lower speed. Irrespective of the biasing conditions, it is desirable that the functionality of the analog compute and memory circuits remain invariant. This property is called bias-scalability~\cite{minggu_thesis}.

Bias-scalable analog circuits were first reported in~\cite{synth_bias_scale} and used a margin-propagation (MP) principle~\cite{mp_patent}. In~\cite{theory_sac}, the MP principle was generalized to shape-based computing, which endowed the analog computing circuits to be process-scalable and temperature-scalable. This work proposes a near-memory shape-based analog computing (S-AC) processing core that combines S-AC multipliers, S-AC non-linear activation, and S-AC compressive digital-to-analog converter (DAC), which are integrated in proximity with each other and are important for ML applications. We have exploited the bias and temperature scalability of the MP principle to achieve faster training (using above threshold biasing) and achieve energy-efficient inference (using sub-threshold biasing) without resorting to any post-training calibration. All the results reported in this paper have been measured from a fabricated prototype, unlike our previous results reported in~\cite{theory_sac} which were based on  circuit simulation of basic computational units. As a proof of concept, this paper also presents the standard regression task at variable speed and power consumption.

The key contributions of this work in relation to previous approaches are as follows:
\begin{itemize}
    \item Design proposition of S-AC based near-memory analog compute core as depicted in Fig.~\ref{blockML_b} which utilizes S-AC multiplier, S-AC non-linearity, and S-AC log-compressive memory digital-to-analog converter (DAC) at its core.
    \item Design proposition of digitally programmable compressive memory DAC utilized as a transformation block near the processing element and at the interface to the external world. This enables scalable memory elements along with scalable analog ML core for faster system adaptation.
    \item Design implementation of S-AC based four quadrant approximate multiplier.
    \item Design validation of proposed S-AC core to implement a standard ML regression task. This results in a near-memory S-AC compute paradigm where the designed S-AC analog ML system can be scaled both for performance or energy. We also verify the functional capabilities of the S-AC system to operate at different computational accuracy and power while maintaining the overall system's performance.
\end{itemize}

\begin{figure*}[t]
	\centering
	\subfloat[]{\includegraphics[width=0.45\linewidth]{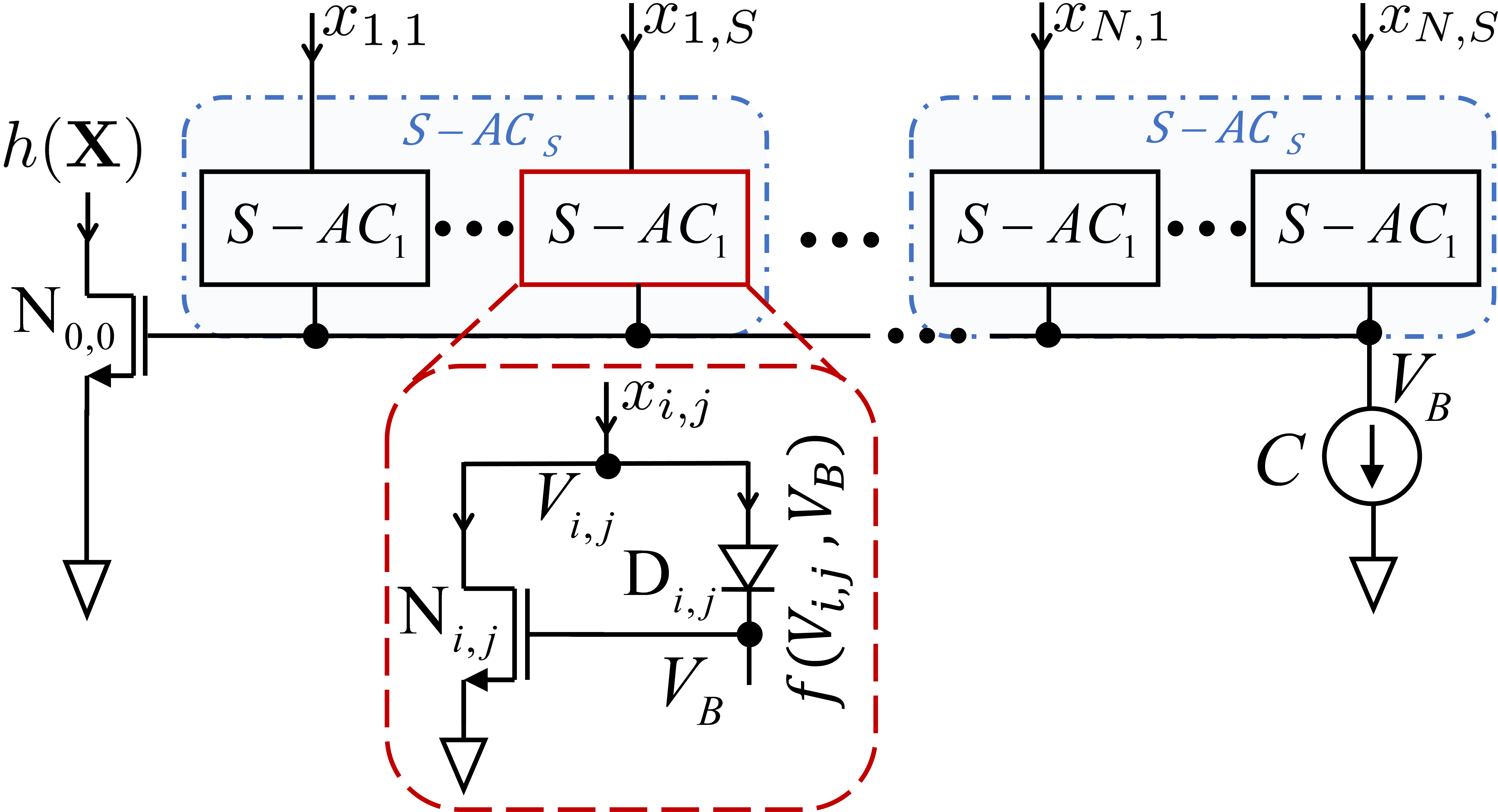}
		\label{sbac_unit}}
	\subfloat[]{\includegraphics[width=0.283\linewidth]{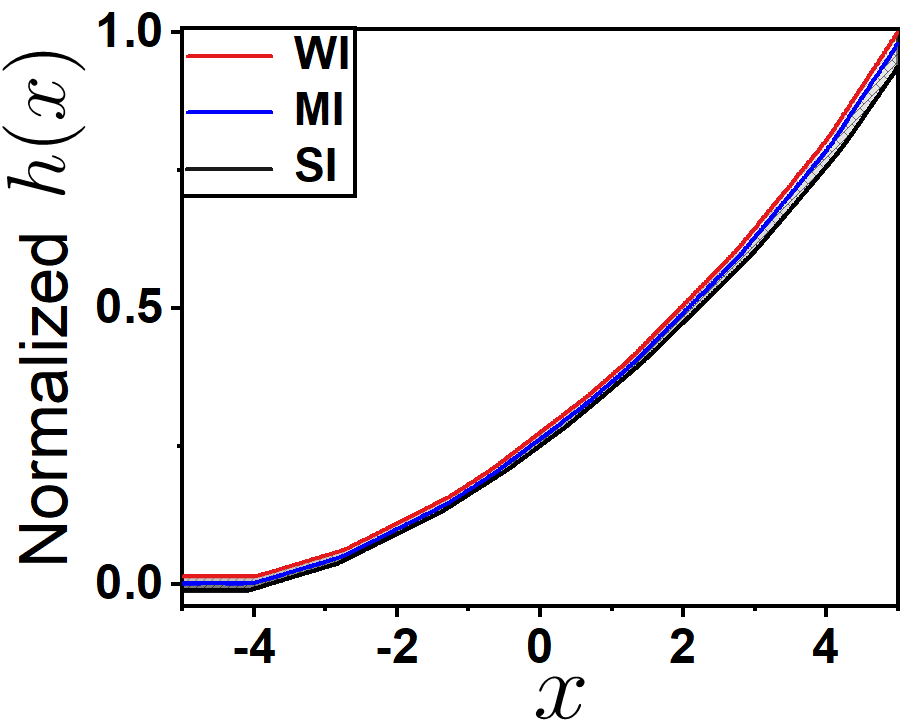}
        \label{shape1_4spline}}
	\subfloat[]{\includegraphics[width=0.268\linewidth]{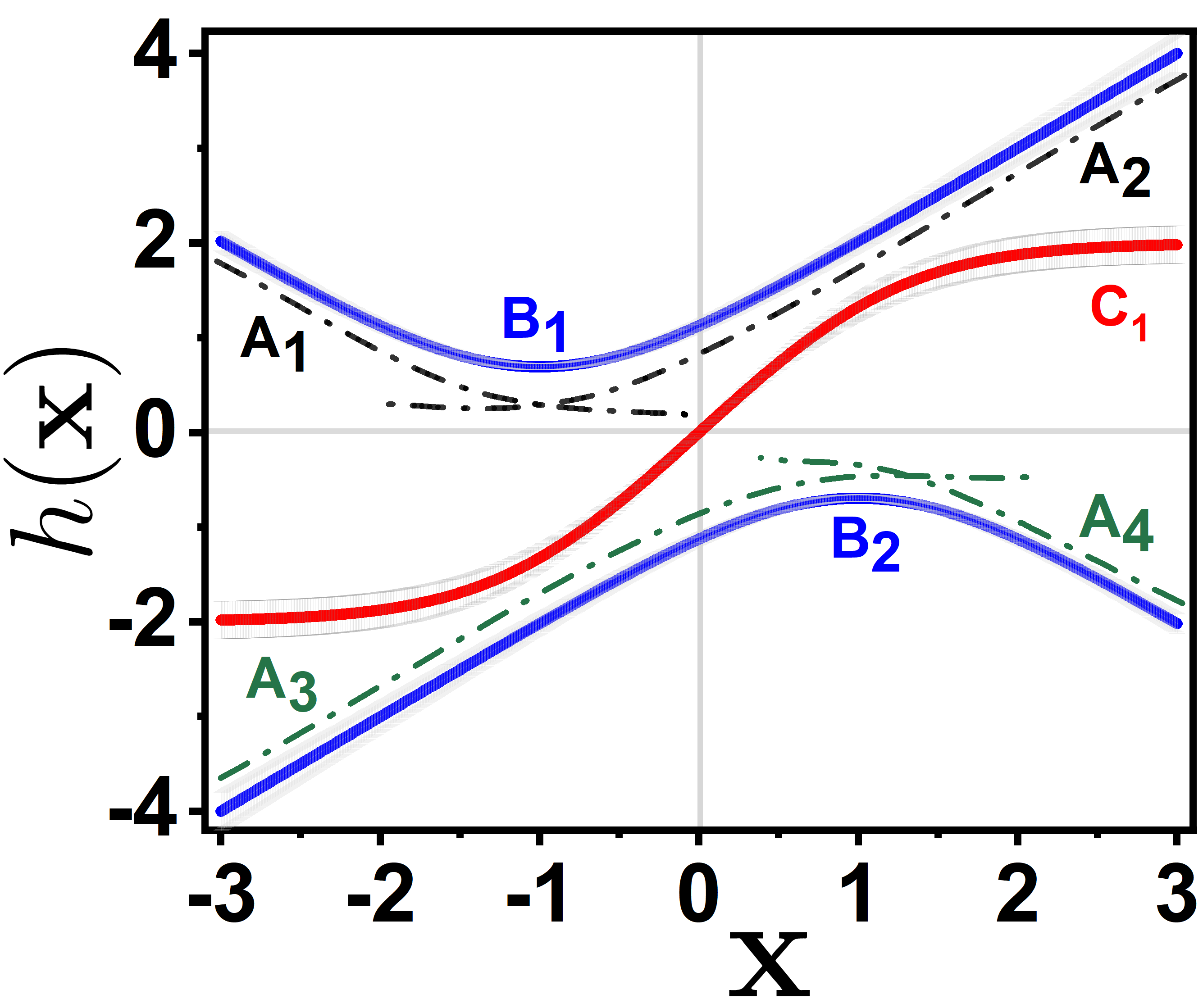}
		\label{multiCurve}}
	\caption{Background on S-AC and MP circuits: \protect\subref{sbac_unit}~Implementation of N-type S-AC circuit~\cite{synth_bias_scale, theory_sac} for $N$ inputs and $S$ splines, the inset shows the circuit implementation of a single S-AC unit using n-type FET and a diode; \protect\subref{shape1_4spline}~Basic S-AC shape (also called proto-shape) implemented by the N-type S-AC circuit when a single input $x$ is varied for different operating regimes and design parameter $S = 3$; \protect\subref{multiCurve}~Different non-linear functions $({A_1} - {A_4}), ({B_1} - {B_2})$ that can be implemented by translation, rotation, addition and subtraction of the proto-shape shown in Fig.~\ref{shape1_4spline}.
	}
	\label{shape_sim}
\end{figure*}

The rest of the sections are organized as follows. Section~\ref{sac_text} describes the background of S-AC circuits and GMP formulation. Section~\ref{design_bloc} shows the design of basic computational blocks for the S-AC processor. Section~\ref{sim_meas} presents the measurement results using prototypes fabricated in the 180nm standard CMOS process, and Section~\ref{Design_Exp} presents the design space analysis of S-AC designs. Section~\ref{appl} demonstrates a simple regression task combining the basic S-AC circuits utilizing an S-AC analog ML core. Finally, in Section~\ref{conc}, we conclude the paper with brief discussions and a comparison with other related work.    

\section{Background on S-AC Circuits and GMP Formulation}\label{sac_text}

In this section, we briefly describe the S-AC approach and generalized margin-propagation (GMP) formulation~\cite{theory_sac,synth_bias_scale} along with its corresponding circuit. The goal of this formulation is to create a robust non-linear shape (also called proto-shape) that depends only on the generic properties of transistors and remains invariant to biasing conditions and operating temperatures. One such methodology to create a robust non-linear shape is described in~\cite{synth_bias_scale,theory_sac}. The function creating this shape is given by
\begin{equation}
    \sum\limits_{i = 1}^N {\sum\limits_{j = 1}^S f ({V_{i,j}},{V_B})}  = C, \forall i = 1,..,N, \forall j = 1,..,S
    \label{char1_msmp}
\end{equation}
\begin{equation}
    f({V_B},0) - f({V_B},{V_{i,j}}) + f({V_{i,j}},{V_B}) = {x_{i,j}}
    \label{char2_msmp}
\end{equation}
where $f:\mathbb{R}\times \mathbb{R} \rightarrow \mathbb{R}$ is a function that models the forward and reverse currents with respect to the gate ($V_g$), drain ($V_d$) and source ($V_s$) voltages respectively of a MOSFET~\cite{tsividis_mos}, $C$ is a hyper-parameter, $x_{i,j}$ is the input, $V_{i,j}$ is an internal variable and $V_B$ is the solution to~\eqref{char1_msmp} and \eqref{char2_msmp}. It can further be re-iterated from~\cite{theory_sac} that \eqref{char1_msmp} is a non-linear constrained optimization problem having a unique solution given by $f(V_B)$. An observation of~\eqref{char1_msmp} and~\eqref{char2_msmp} also reveals that if the source and drain terminals are symmetric to each other, the expressions in~\eqref{char1_msmp} and~\eqref{char2_msmp} are true despite the choice of transistor models (EKV~\cite{ekv}, ACM~\cite{acm}, etc.) or operating regimes, (weak-inversion WI, moderate-inversion MI or strong-inversion SI), or process nodes (MOSFET, FinFET, etc.). This makes S-AC circuits truly scalable across different bias conditions.

Now for a given input matrix $X\in\mathbb{R}^{N\times S}$ where $\textbf{x}_{i} \in \mathbb{R}^{N}$, $\forall i = 1,..,N$ is the input vector and $\textbf{x}_{j}\in \mathbb{R}^{S}$, $\forall j = 1,..,S$ is the number of splines implementing the approximation, the CMOS circuit satisfying constraints in~\eqref{char1_msmp} and \eqref{char2_msmp} is shown in Fig.~\ref{sbac_unit}. Here, $x_{i,j}$ is the input current for the $i^{th}$ input and the $j^{th}$ spline, $h(X) = f(V_B,0)$ is the output current, $V_{i,j}$ and $V_B$ are the voltages across the $N_{i,j}^{th}$ transistor, $C$ is a constant current, and $D_{i,j}$ denotes diode elements (Schottky, MOS diode or any other). Applying KCL at node $V_B$,~\eqref{char1_msmp} can be obtained while the current across diode $D_{i,j}$ gives~\eqref{char2_msmp}. 

It can further be noted that the current source $C$ implements a constraint that dictates all the diode currents to add upto $C$ where the diode element itself forces the flow of current in one direction, similar to rectification operation. In addition, transistors  $N_{i,j},~\forall i = 1,..,N$ and $j = 1,..,S$ form the current mirror branches, thereby enforcing similar currents among all the mirror branches. Due to these simultaneous constraints imposed by the current source $C$, diode elements $D_{i,j}$, and the mirror branches $N_{i,j}$, the implemented circuit in Fig.~\ref{sbac_unit} settles to a unique value of $V_B$  which satisfies all these constraints. This unique value of $V_B$ finally results in output $h$ \textit{i.e.} $h(X) = f(V_B,0)$ as the solution to \eqref{char1_msmp} and \eqref{char2_msmp}. Furthermore it can be emphasized from~\cite{theory_sac} that the output $h(\cdot)$ (also called proto-shape) similar to function $f(\cdot)$ always satisfies the properties \eqref{eq:lipschitz} and \eqref{eq:asymptote2} given by 
\begin{equation}
    1 \ge \frac{\partial h}{\partial x_i} \ge 0, \forall i
    \label{eq:lipschitz}
\end{equation}
\begin{align}
	\begin{split}
\mathop {\lim }\limits_{{x_i} \to \infty } \frac{{\partial h}}{{\partial {x_i}}} = 1\\
\mathop {\lim }\limits_{{x_i} \to  - \infty } \frac{{\partial h}}{{\partial {x_i}}} = 0
	\end{split}
	\label{eq:asymptote2}
\end{align}
The property in~\eqref{eq:lipschitz} ensures that the obtained proto-shape $h$ is monotonic with respect to its variable, while the properties described by \eqref{eq:asymptote2} determine the two asymptotes of $h$, irrespective of the specific form of $f$. Fig.~\ref{shape1_4spline} shows the example of the obtained shape using the circuit in Fig.~\ref{sbac_unit} for input dimension $N=1$ and the design parameter $S=3$. The results are also shown for different MOSFET biasing regimes, i.e., WI, MI, and SI biasing regimes which correspond to different functions $f(\cdot)$ in~\eqref{char1_msmp} and \eqref{char2_msmp}. This shape will often be referred to as the basic proto-shape in the following text. Fig.~\ref{multiCurve} shows that other complex non-linear monotonic shapes can further be constructed using the proto-shape by using only rotation and translation techniques.
 
\begin{figure*}[t]
	\centering
	\subfloat[]{\includegraphics[width=0.635\linewidth]{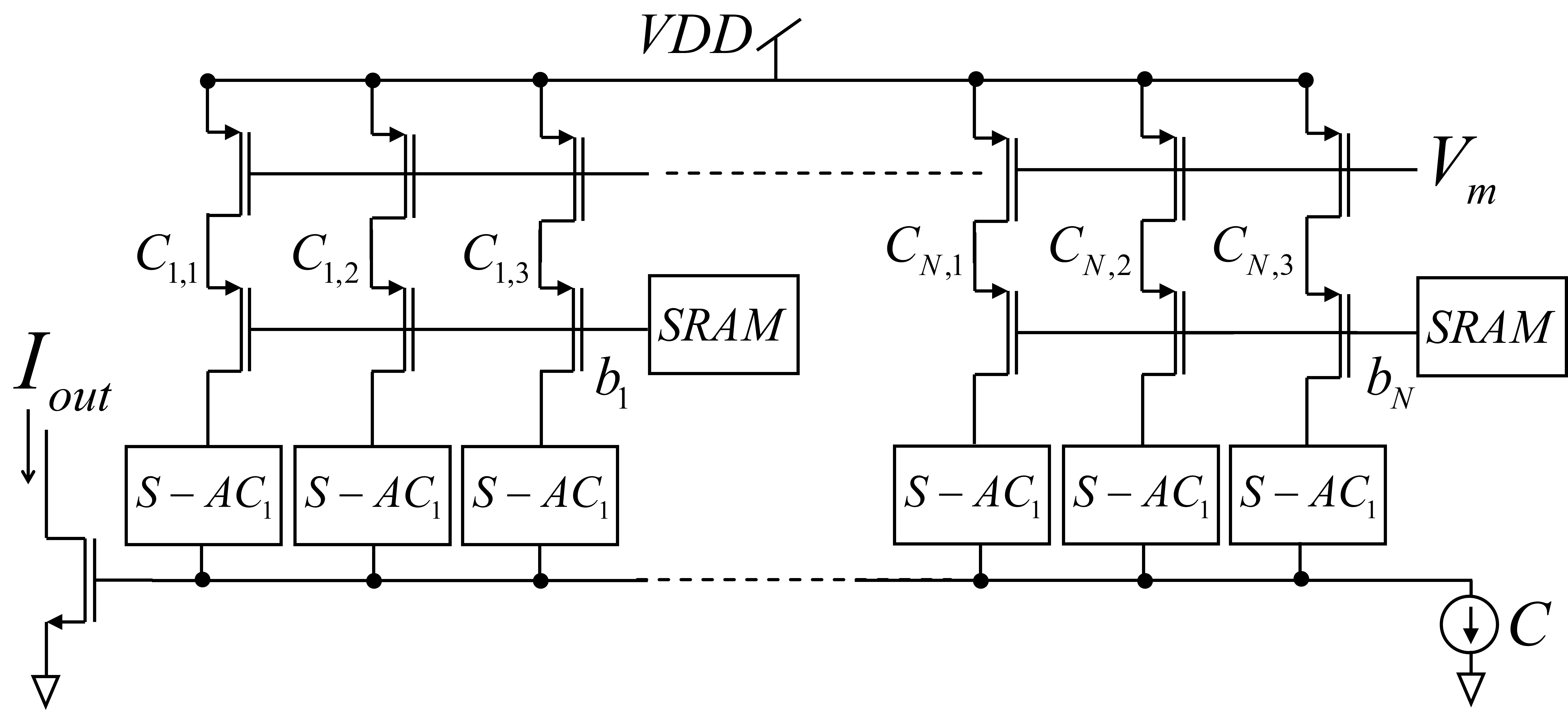}
	\label{DAC_ckt}}
	\hspace{0.2 cm}
	\subfloat[]{\includegraphics[width=0.3\linewidth]{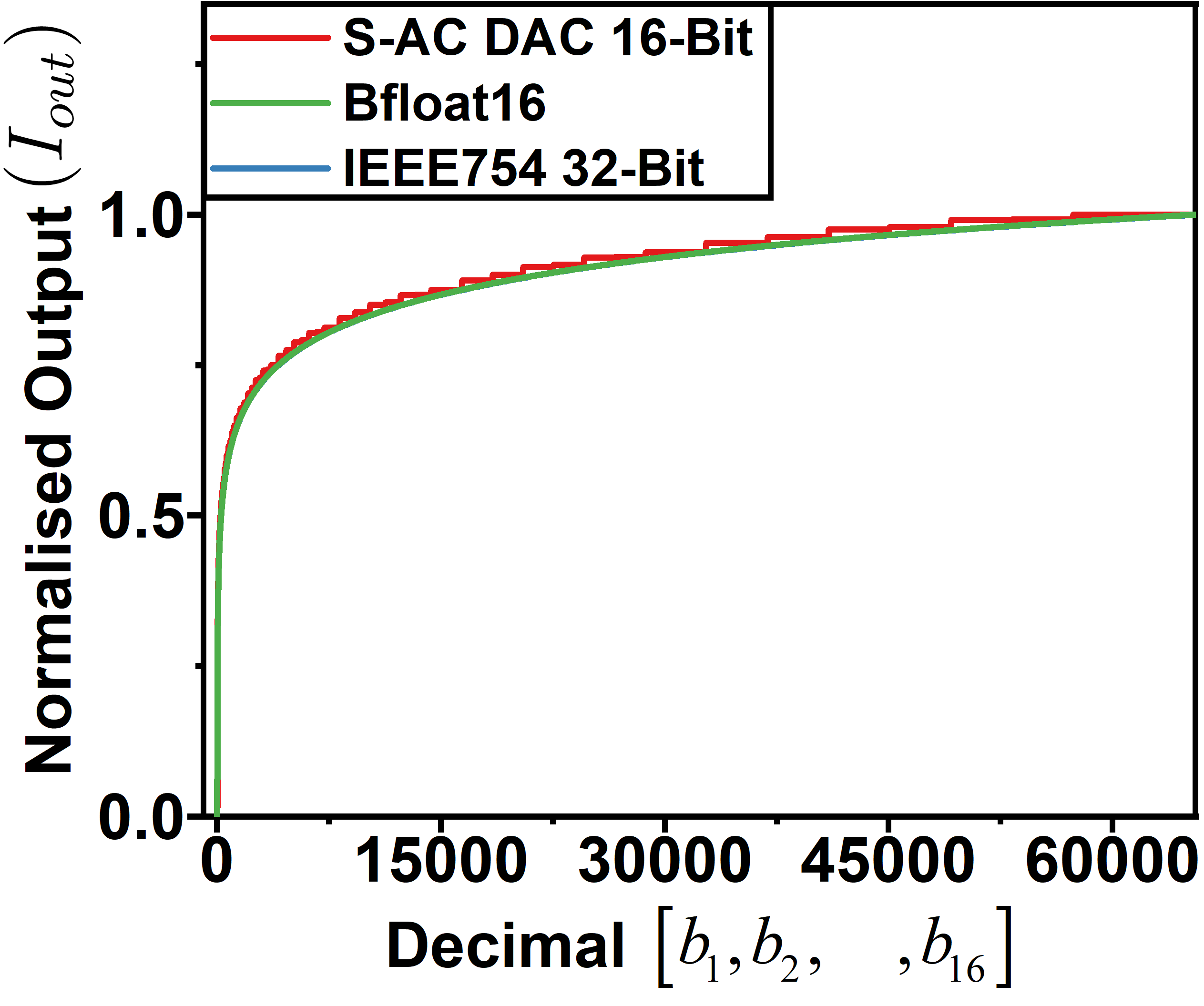}
	\label{bfloat16}}
	\caption{ \protect\subref{DAC_ckt} Compressive log-binary DAC implementation using S-AC; \protect\subref{bfloat16} Comparison plot between S-AC log-binary DAC, Bfloat16 \& IEEE32 number systems demonstrating close compliance of compressive nature of corresponding $log_2$ curves with each other.}
\end{figure*}

 \section{Near-Memory S-AC Core Design}\label{design_bloc}

This section presents the building blocks required for designing a near-memory analog ML core. An adequate requirement for an ML inference processor incorporating analog ML cores requires: (a) memory for storing the inference parameters and for supporting a digital interface for inputs; (b) multiply-accumulate circuits; and (c) non-linear computing circuits. Here we show that the basic S-AC circuit shown in Fig.~\ref{sbac_unit} can be modified and extended to implement all the required computational building blocks. We specifically implement a combination of a compressive mixed-signal memory DAC and a non-linear multiplier circuit that results in a multiply-accumulate (MAC) operation, which also emulates computing using Bfloat16 number representation~\cite{bfloat_cloud_tpus}. It may be noted that any approximation error introduced in this mapping can be compensated during training itself, as neural networks are resilient to error and can adapt to approximation errors if properly trained~\cite{multiplier_abhishek}.
 
 \subsection{S-AC based Compressive Memory DAC}\label{dac_text}

One of the major challenges in implementing an analog ML processor is storing and updating trained parameters. While analog memories based on memristors, floating gates, and other nano-scale devices have been proposed for analog ML processors~\cite{Xia20,Seb20,Bay17}, their functional response and speed do not scale across training and inference. Therefore, in this paper, we propose to use a DAC-based memory that uses a S-AC based analog front-end to implement a compressive function. This compressive function will be utilized by the S-AC multiplier (discussed in Section~\ref{mult_text}) to implement a memory-compute block for MAC operation. Here we show that this compressive-expansive operation is equivalent to analog computing using Bfloat16~\cite{bfloat_cloud_tpus} and the IEEE-754 single-precision (32-bit) number systems. Note that the Bfloat16 number system developed by Google Brain delivers more accurate results at lesser hardware as compared to IEEE 754 single-precision numbers for some neural networks and is extensively used by Google cloud TPUs~\cite{bfloat_cloud_tpus}.
Consider a function $g(\bf{x})$ given by
\begin{equation}
 g\left( {\bf{x}} \right) = {\log _2}\left( {\sum\limits_i {{2^{{x_i}}}} } \right).  
\label{genericLogEq}
\end{equation}
Then, it is easy to verify that $g(\bf{x})$ satisfies the properties 
\begin{equation}
  1 \ge \left| {\frac{{\partial g}}{{\partial {x_i}}}} \right| \ge 0
  \label{lip1}
\end{equation}
\begin{equation}
  \mathop {\lim }\limits_{{x_i} \to \infty } \frac{{\partial g}}{{\partial {x_i}}} = 1 
  \label{lip2}
\end{equation}
similar to that of the proto-shape $ h\left( {\bf{x}} \right)$ in \eqref{eq:lipschitz} and \eqref{eq:asymptote2}. 
If $\bf{\x}$ is denoted by its binary representation as $\bf{\x}$ $\cong \sum\limits_{i = 1}^N 2^i b_i$, then incrementing the design parameter $S$ per bit, we have
\begin{equation}
g\left( \mathbf{x} \right)={{\log }_{2}}\sum\limits_{i=1}^{N}{\sum\limits_{j=1}^{S}{{{2}^{{{C}_{ij}}}}{{b}_{i}}}}={{\log }_{2}}\sum\limits_{i,j:{{b}_{i}}=1}^{NS}{{{2}^{{{C}_{ij}}}}}=g\left( \mathbf{B} \right)
\label{genericdac_2}
\end{equation}
where $\mathbf{B}$ $\in\{0,1\}^S\times\{0,1\}^N$ is a binary input matrix and $N$ is the number of inputs. It can be seen that~\eqref{genericdac_2} (logarithmic DAC) is a special case of~\eqref{genericLogEq} and hence can be approximated using the proto-shape $ h\left( {\bf{x}} \right)$. 
Fig.~\ref{DAC_ckt} shows the circuit implementation of N-bit S-AC based compressive memory for $S=3$. Switches connected at ${b_1}, {b_2}$,...., $b_N$ are implemented using transmission gate (TG) switches. Here,  $[b_1,....,b_{N-1}, b_N]$ represent an N-bit binary number to be converted into its analog equivalent and ${[C_{1,1}},{C_{1,2}},{C_{1,3}},...,{C_{N,3}]}$ are the offsets when $S=3$ as can be calculated from~\cite{theory_sac}. The proposed S-AC based DAC converts the digital input into a compressive analog output. It may be noted that this compressive output is implicitly expanded in~\eqref{ip_expanded} for multiplication. 

We also extend this analysis to implement any generic-base DAC. Using the base change property of logarithm, \eqref{genericdac_2} can be rewritten for any base $\theta$ as
\begin{align}
g\left( {\bf{x}} \right) = \frac{{g\left( {\bf{B}} \right)}}{{{{\log }_\theta }\left( 2 \right)}}
\label{genericdac_6}
\end{align}
This implies that by scaling the output by a factor ${{{\log }_\theta }\left( 2 \right)}$, any \textit{log-Generic-DAC} can be implemented from an existing DAC using \textit{S-AC} framework. Fig.~\ref{bfloat16} compares the $log_2$ characteristics of the \textit{Brain float} (Bfloat16) and the IEEE-754 single-precision (32-bit) number systems for 16-bit numbers normalized between 0 to 1, and the response obtained using the S-AC DAC for design parameter $S=4$. The results show compliance between the different logarithmic number representations. It can be noted that the goal here is not to mimic the exact Bfloat16 response but to get a compressive response that matches the required number system. Rest can be assumed to be calibrated while training. 

 \begin{figure*}[t]
	\centering
	\subfloat[]{\includegraphics[width=0.55\linewidth]{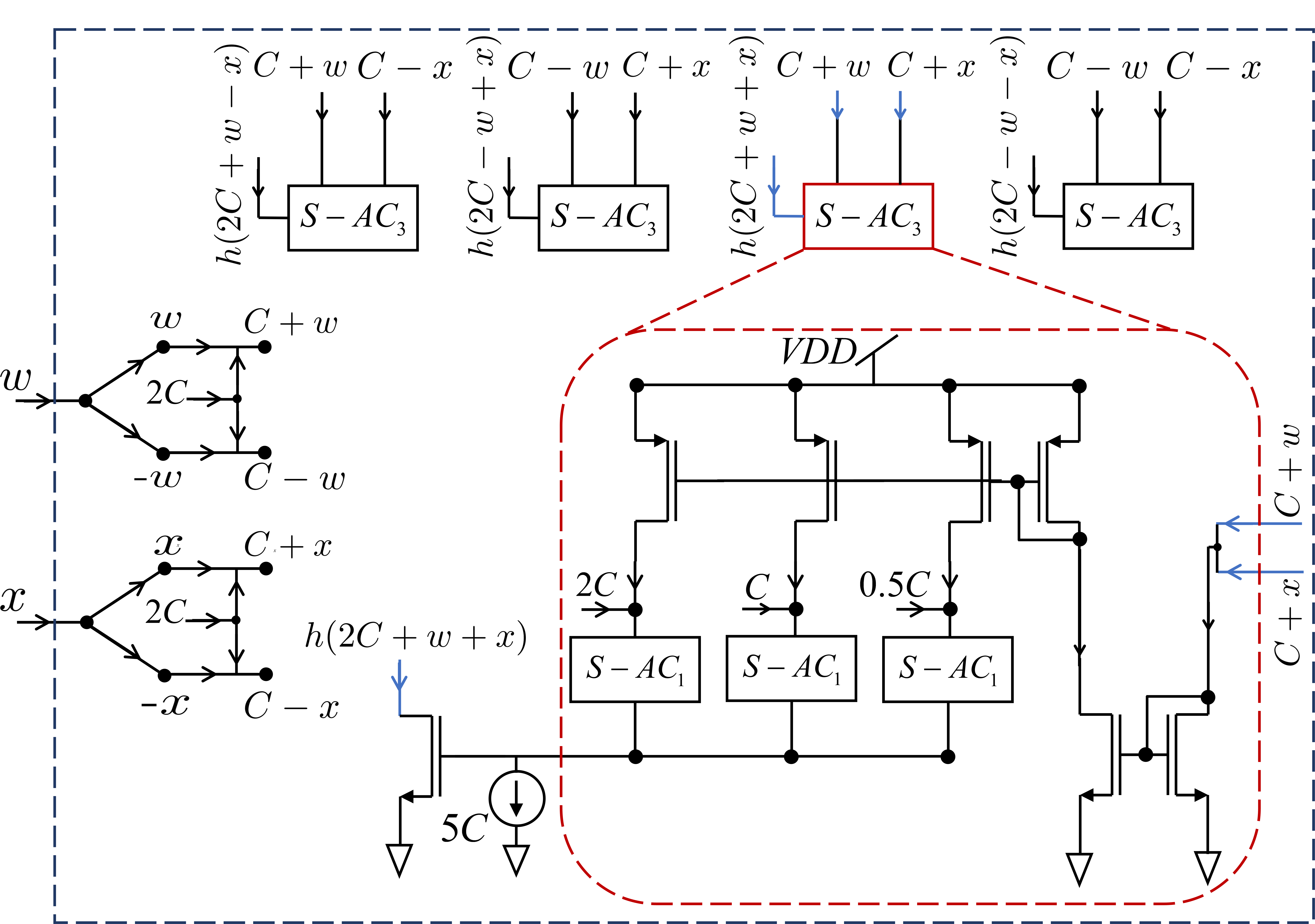}
	\label{multiply_ckt}}
	\subfloat[]{\includegraphics[width=0.35\linewidth]{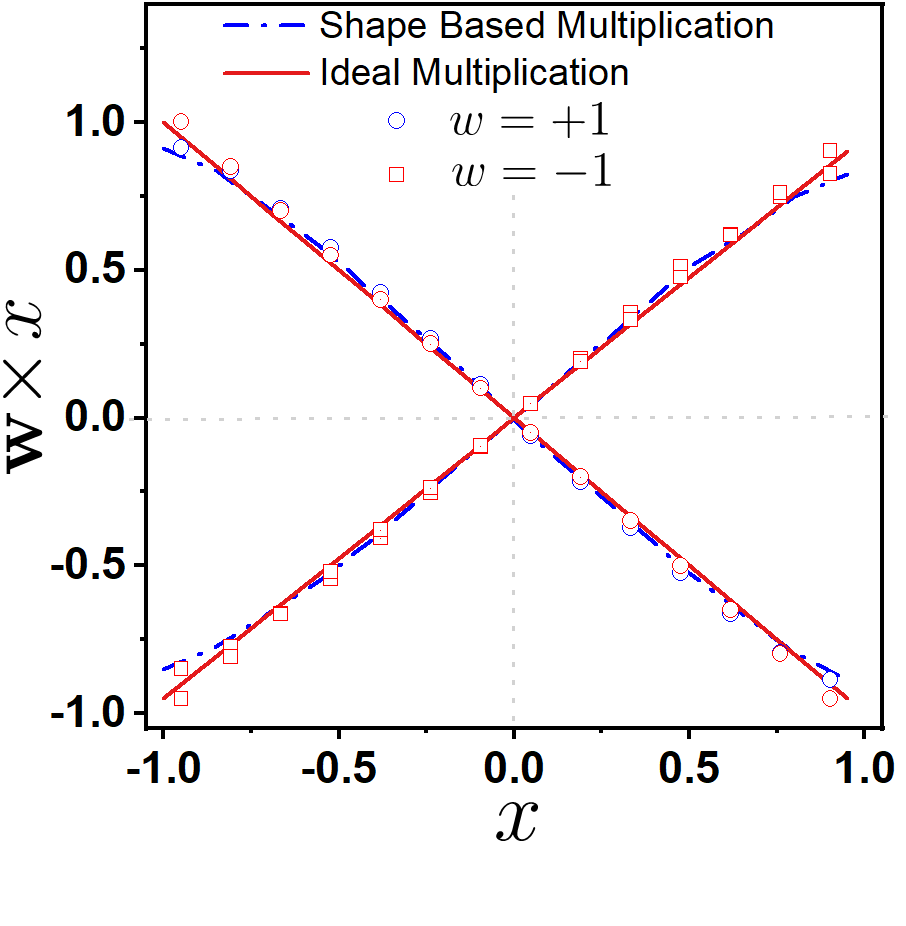}
	\label{scalar_innerpro}}
	\caption{ \protect\subref{multiply_ckt} Implementation of S-AC multiplier for design parameter $S=3$; \protect\subref{scalar_innerpro}  Comparison of four-quadrant S-AC multiplication with ideal multiplication.}
\end{figure*}

\subsection{S-AC based Analog Multiplier}\label{mult_text}

The S-AC proto-shape $h$ can be used to implement analog multipliers based on the following Taylor series approximation
\begin{align}
\begin{array}{l}
h\left( {C + w + C + x} \right) - h\left( {C + w + C - x} \right) \ldots \\
 \quad\quad+ h\left( {C - w + C - x} \right) - h\left( {C - w + C + x} \right)\\
 \quad\quad\approx 2x \times \left( {\frac{{dh\left( {C + w} \right)}}{{dw}} - \frac{{dh\left( {C - w} \right)}}{{dw}}} \right)\\
 \quad\quad\approx 2x \times \left( {{w^ + } - {w^ - }} \right)\\
 \quad\quad\approx 2x \times w
 \label{ip_expanded}
\end{array}\
\end{align}
The constant $C$ ensures that the input to the proto-shape is always positive. The differential combination effectively cancels the zeroth order and second-order terms in the Taylor series~\cite{multiplier_abhishek} and the property of $h$ in~\eqref{eq:lipschitz}, leads to~\eqref{ip_expanded}. The detailed derivation is provided in Appendix~\ref{Appendix:A}. Note that one of the differential arguments to the multiplier $\left( w^+ - w^- \right)$ is a non-linear map $\frac{dh}{d{w}}$, which, based on property~\eqref{eq:lipschitz}, is a compressive map. Thus, the stored parameters need to be pre-processed before and are presented as an input to the multiplier. This is the basis for our compressive memory design described in Section~\ref{dac_text}.   

The circuit in Fig.~\ref{multiply_ckt} implements the scalar multiplication given in~\eqref{ip_expanded} where $w \in \mathbb{R}$, $x \in \mathbb{R}$  and the product $y \in \mathbb{R}$. Fig.~\ref{multiply_ckt} shows the S-AC unit utilized to implement each component in~\eqref{ip_expanded}. The inputs are first converted into  their differential forms, and constant ($C$) is added to the negative term to shift the operation in the first quadrant. The output from all S-AC$_3$ (here subscript $3$ represents $3-$spline S-AC) units is added and subtracted (differentially) as per~\eqref{ip_expanded} to obtain the desired multiplication. Fig.~\ref{scalar_innerpro} shows a close approximation between the simulated output of the four-quadrant multiplier and the output obtained from an ideal multiplier. Based on this basic operation, multiply-accumulate operations and inner products can now be implemented by combining element-wise S-AC multipliers with summing circuits based on Kirchhoff's current law. Other parallel analog matrix-vector-multiplier architectures have been reported in the literature~\cite{mvm1,mvm2}.
\begin{figure}[t]
    \centering
    \includegraphics[width=0.45\linewidth]{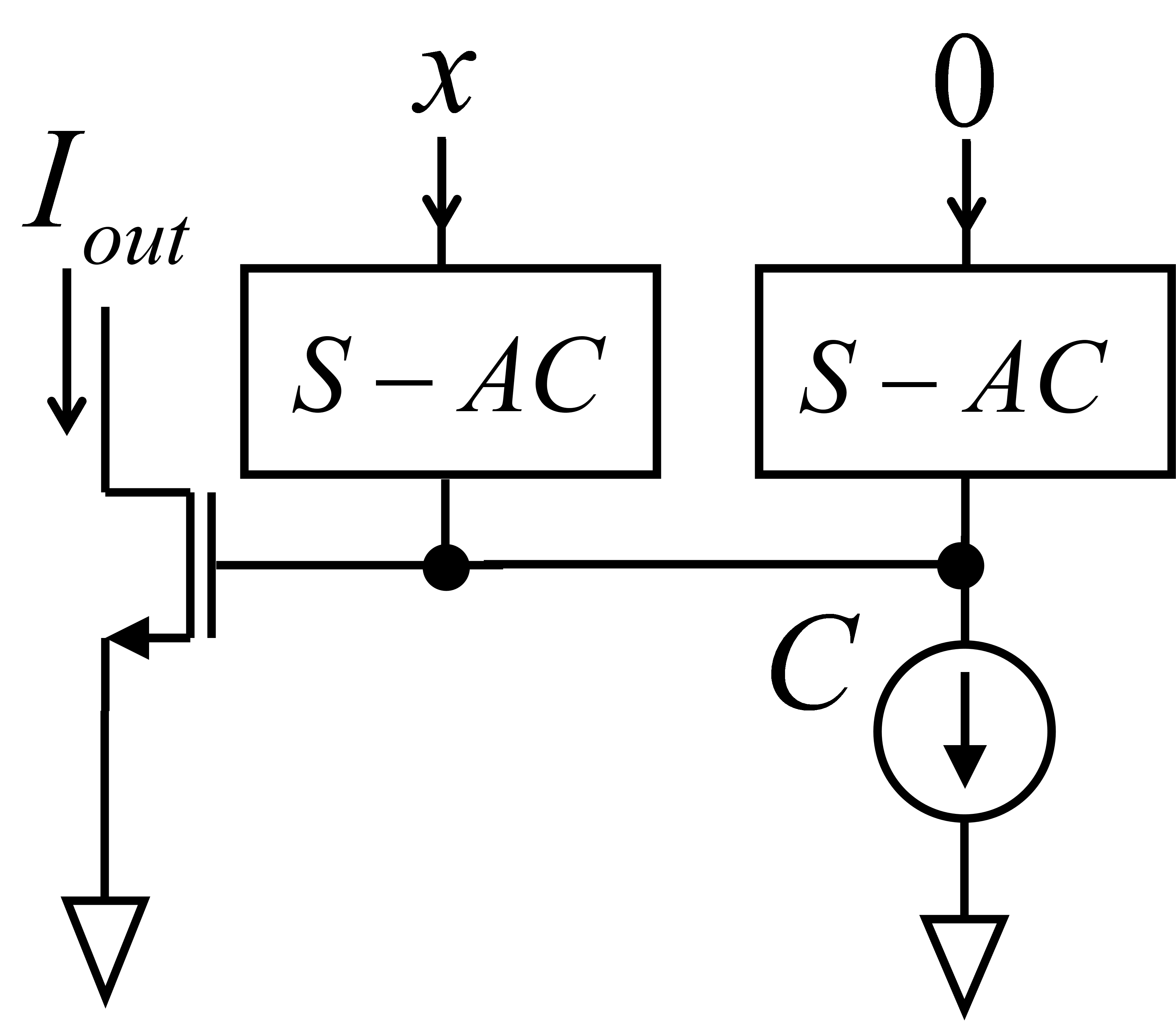}
    \caption{Implementation of Soft ReLU activation using S-AC.}
    \label{nl_ckt}
\end{figure}

 \subsection{S-AC based ReLU Activation}\label{nonlin_text}
A soft ReLU function can be implemented using a one-dimensional proto-shape shown in Fig.~\ref{shape1_4spline}. The circuit implementation of the soft ReLU function is shown in Fig.~\ref{nl_ckt}. The basic circuit uses two S-AC units, one of which receives an input $x$, and the other is driven by a zero current (or floating). It may be noted that as limit $C \rightarrow 0$, the proto-shape converges to an ideal ReLU function. 
Other non-linear functions can also be implemented by shift, translation, and addition of the basic proto-shape, as illustrated in Fig.~\ref{multiCurve}.

\section{Measurement Results}\label{sim_meas}

\begin{figure}[t]
	\centering
	\subfloat[]{\includegraphics[width=0.9\linewidth]{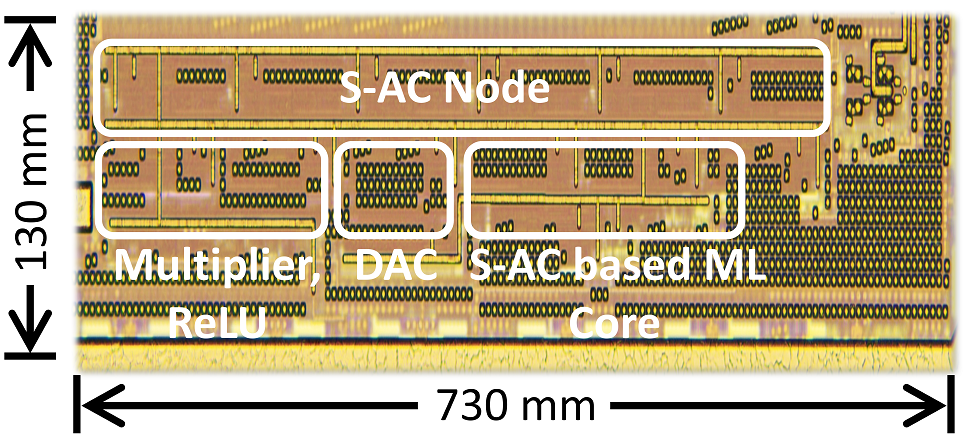}
	\label{chip}}
	
	  \vspace{-3mm}
	\subfloat[]{\includegraphics[width=0.8\linewidth]{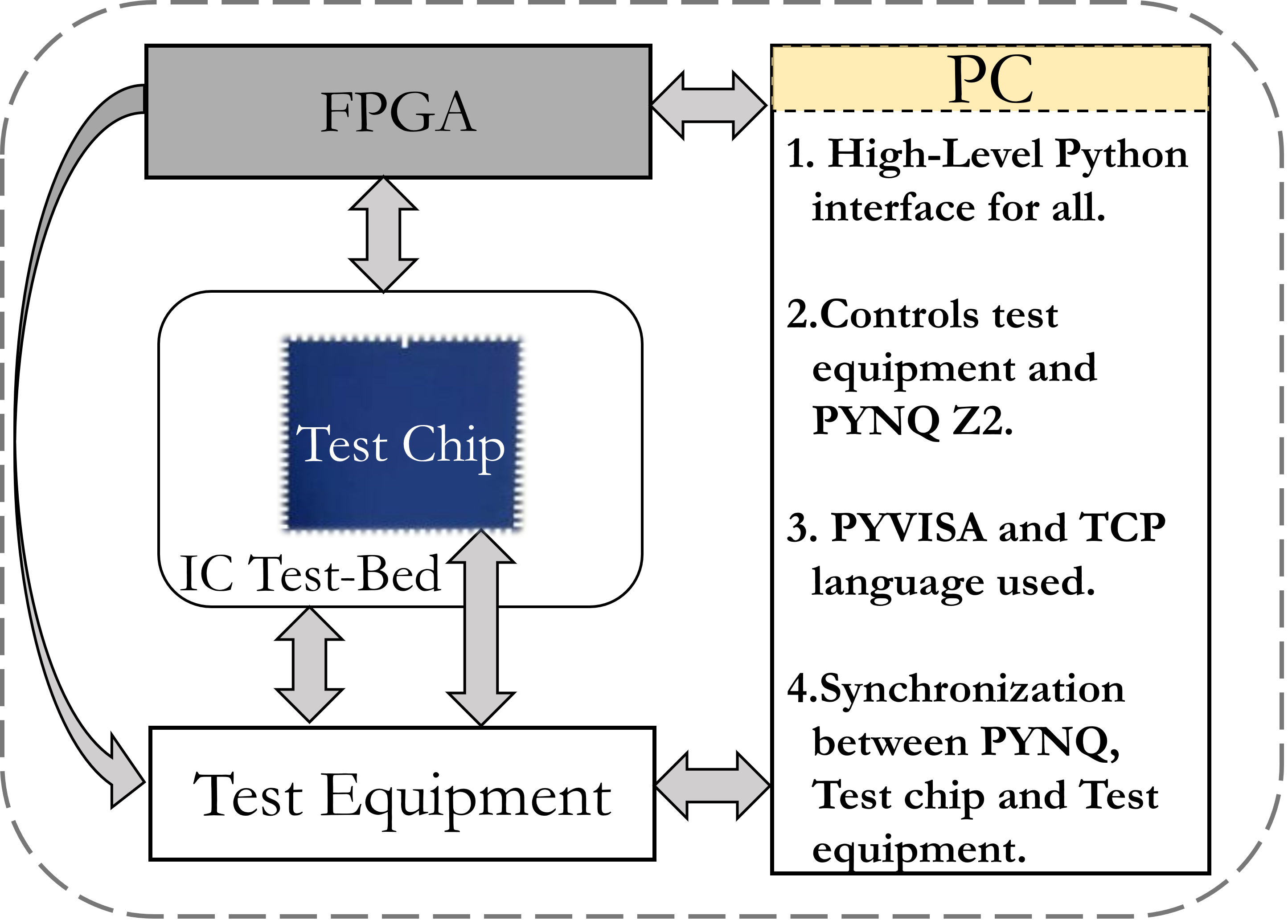}
	\label{test}}
 	\caption{\protect\subref{chip} Die micro-photograph of the chip; \protect\subref{test} Test measurement setup.}
	\label{chip_test}
\end{figure}

The S-AC building blocks, along with computational nodes of S-AC based neural network, have been prototyped in a standard CMOS 180nm process technology. Fig.~\ref{chip} shows the die microphotograph of the chip where a copy of basic computational blocks and the S-AC node has been highlighted. It may be noted that multiple copies of basic computational blocks were fabricated for test purposes. The functionality of the circuit modules has been verified using the test measurement setup shown in Fig.~\ref{test}. The test chip was mounted on a custom IC test board, and the test vectors were generated using a PYNQ-Z2 FPGA board which used a python-based interface to control the digital inputs and outputs. High-precision analog test equipments were directly interfaced with the test chip and were controlled by the PYNQ-Z2 FPGA board. To accurately determine the region of operation, the transistors were first characterized for some fixed circuit parameters (such as the aspect ratio of the transistor, spline count $S$, etc.). The inversion coefficient (IC) range~\cite{binkley2007tradeoffs} was then used to find an approximate one-to-one mapping between the IC range and the bias current range. For strong inversion regime of operation, circuits were biased so as to  maintain the inversion coefficient i.e. $IC>10$ while for weak inversion $IC <0.1$ was maintained and all the way between $0.1<IC<10$ was marked as moderate inversion~\cite{binkley2007tradeoffs}.

\subsection{S-AC Compressive Memory Measured Result}

Fig.~\ref{dac_hw} shows the measured result of 8-bit S-AC based DAC as a function of equivalent decimal input varying from 0 to 255 at different operating regimes. It can be seen that the result closely approximates the desired ideal shape and the output shape is invariant across operating regimes.
With the increase in constant current $C$, along with the offsets ${[C_{1,1}},{C_{1,2}},{C_{1,3}},...,{C_{N,3}]}$ for $S=3$, the S-AC based DAC operation moves from WI to SI resulting in increased power consumption but simultaneously reducing settling time and in turn improving throughput and speed. However, the optimum trade-off between energy and throughput can be obtained in the MI region of operation.

\subsection{S-AC Multiplier Measured Results}

\begin{figure*}[t]
	\centering
	\subfloat[]{\includegraphics[width=0.254\linewidth]{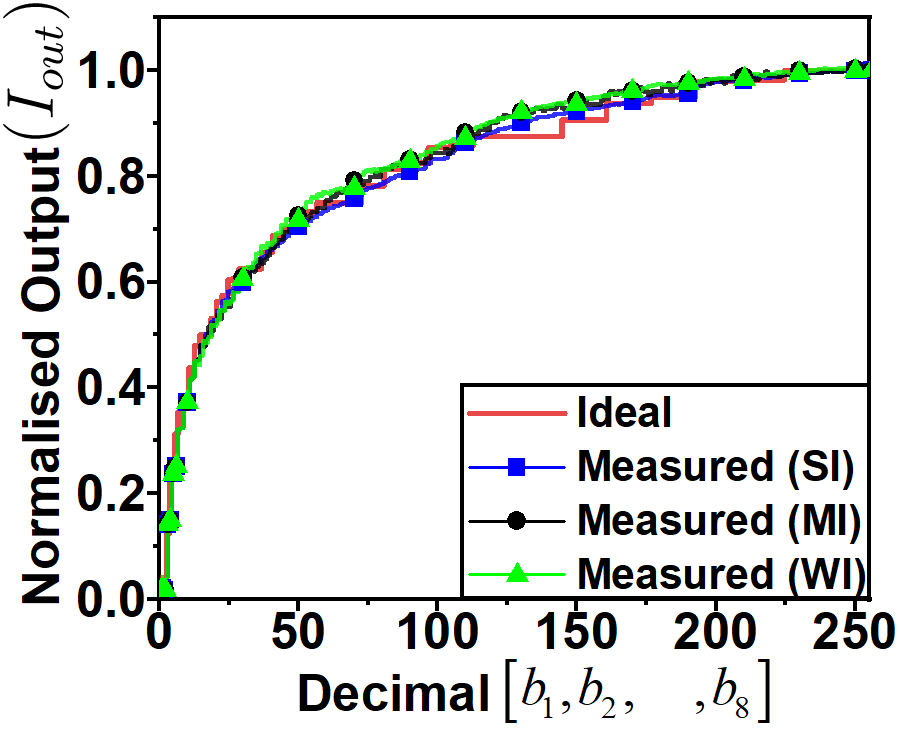}
	\label{dac_hw}}
	\subfloat[]{\includegraphics[width=0.254\linewidth]{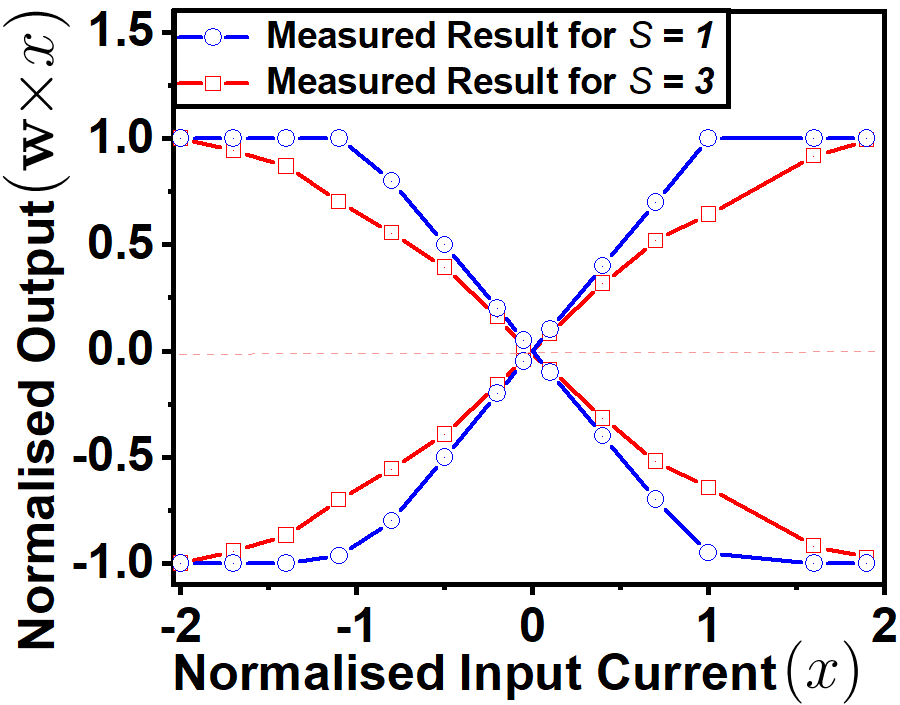}
	\label{multiply_hw1}}
	\subfloat[]{\includegraphics[width=0.249\linewidth]{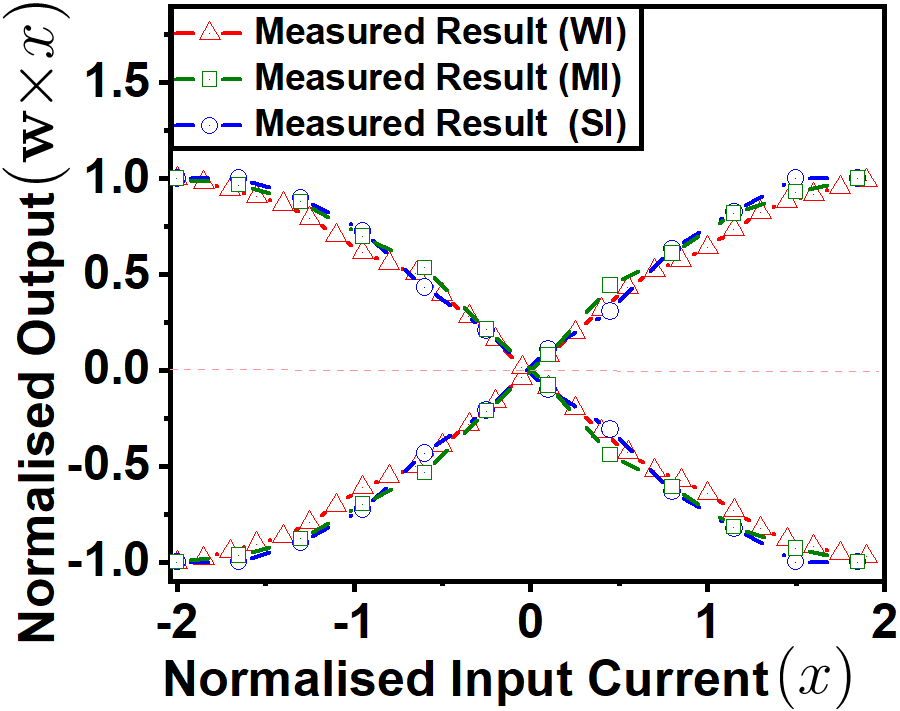}
	\label{multiply_hw2}}
	\subfloat[]{\includegraphics[width=0.253\linewidth]{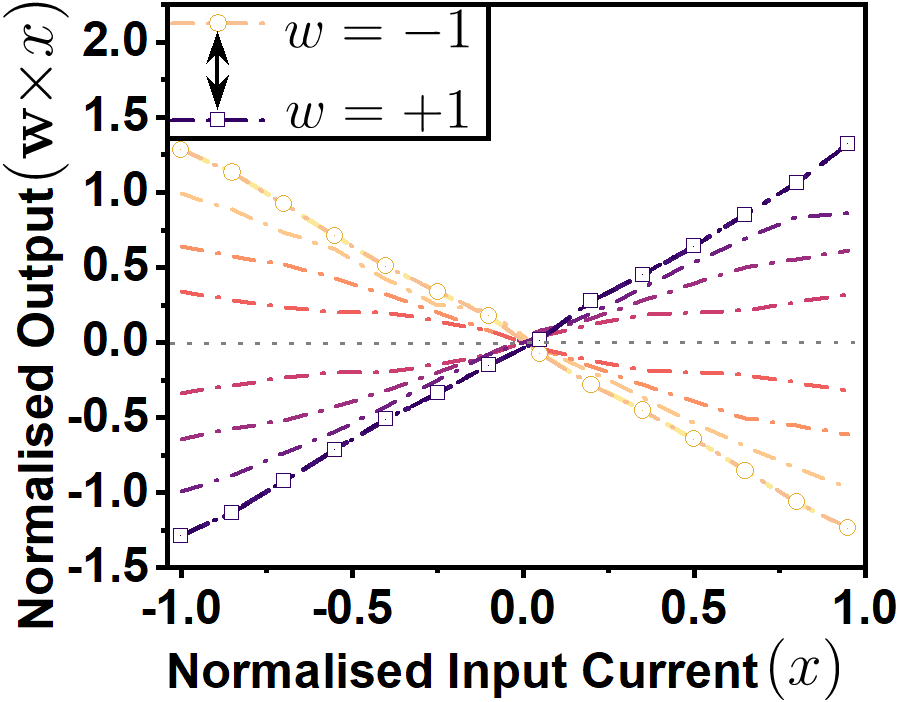}
	\label{multiply_hw3}}
 	\caption{\protect\subref{dac_hw}~Measurement result (normalized) of 8-bit compressive DAC shown in Fig.~\ref{DAC_ckt}; Measurement result (normalized) of four-quadrant S-AC multiplication shown in Fig.~\ref{multiply_ckt} showing: \protect\subref{multiply_hw1}~varying accuracies at different design parameter $S=1$ and $S=3$ for $\bf{w}$$=[- 0.5,0.5]$; \protect\subref{multiply_hw2}~close compliance between multiplier curves at different operating regimes for $S=3$; \protect\subref{multiply_hw3}~multiplier curves for ${\bf w} = [-1, -0.75, -0.5, -0.25, 0.25, 0.5, 0.75, 1]$ at $S=3$.}
	\label{multiplier_res}
\end{figure*}

Fig.~{\ref{multiply_hw1}} and Fig.~{\ref{multiply_hw2}} shows the measured result of the implemented S-AC multiplier circuit for different values of design parameter $S$ and at different operating conditions. Fig.~\ref{multiply_hw1} shows the comparison plot of a four-quadrant multiplier for design parameters $S=1$ and $S=3$.  It can be noted that with the increase in design parameter $S$, the multiplier accuracy increases and becomes much closer to the ideal. Fig.~\ref{multiply_hw2} is computed for $S=3$, and shows the four-quadrant multiplication at different operating regimes in close compliance with each other. The results of Fig.~\ref{multiply_hw1} and Fig.~\ref{multiply_hw2} have been computed for $x \in (- 1,1)$ and for $\bf w$ $=$ $[- 0.5,0.5]$. Fig.~\ref{multiply_hw3} shows the multiplication curve for $x \in \left( { - 2,2} \right)$ and for different values of $w$. 

\begin{figure}[t]
\centering
\includegraphics[width=0.6\linewidth]{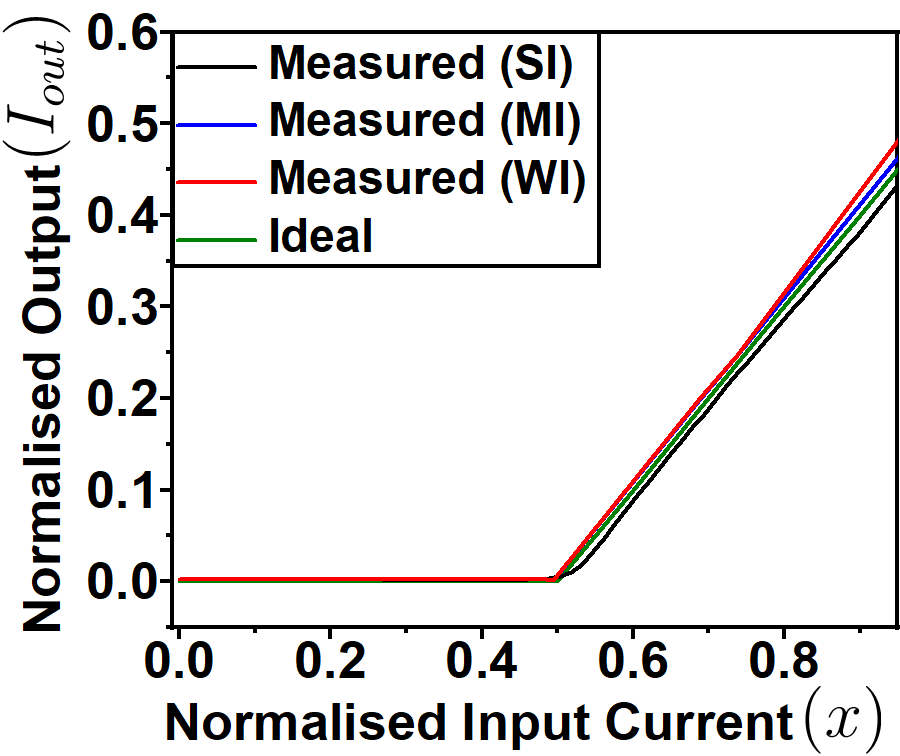}
\caption{Measurement result of S-AC ReLU implementation shown in Fig.~\ref{nl_ckt} for $C=0.5$ demonstrating shape invariance across operating regions.}
\label{relu_hw}
\end{figure}

\subsection{S-AC ReLU Measured Results}

Fig.~\ref{relu_hw} shows the measured results of S-AC based ReLU implementation (Fig.~\ref{nl_ckt}) and its comparison with the ideal. It can be observed that the obtained normalized output current curve follows the desired non-linear shape and matches the ideal. Furthermore, the non-linear shape remains invariant in weak, moderate, and strong inversion regimes as desired. The average power consumption varies between $18.2~nW$ to $89.4~\mu W$ with an area consumption of $190.46$~${\mu m^2}$ in 180nm technology node when the circuit operation shifts from WI to MI, respectively. This is much lesser than the corresponding digital implementation where the power reported in~\cite{Relu_pow} is of the order of $1~mW$ at 40nm technology node with an area consumption of $1697.37$~${\mu m^2}$.

\section{Design Space Trade-Offs Analysis}\label{Design_Exp}

\subsection{S-AC Area and Power Saving Analysis}
S-AC design offers a range of trade-offs between accuracy, area, and power benefits by changing the design parameter ($S$). The value of this design parameter $S$ is determined based on the application requirements. Theoretically, the number of splines selected can vary from $1$ to $S$, therefore offering a wide range of trade-offs to choose from. On analysis, it was found that each increment in design parameter $S$, decreases the approximation error exponentially at the cost of additional $\approx 20\%$ area requirement as compared to previous $S$. The performance evaluation of the S-AC multiplier found that with an average absolute error of 3.66\%, $S = 3$ offers up to 31.3\% in area savings and up to 37.2\% in power savings.
\begin{figure*}[th]
	\centering
	\subfloat[]{\includegraphics[width=0.305\linewidth]{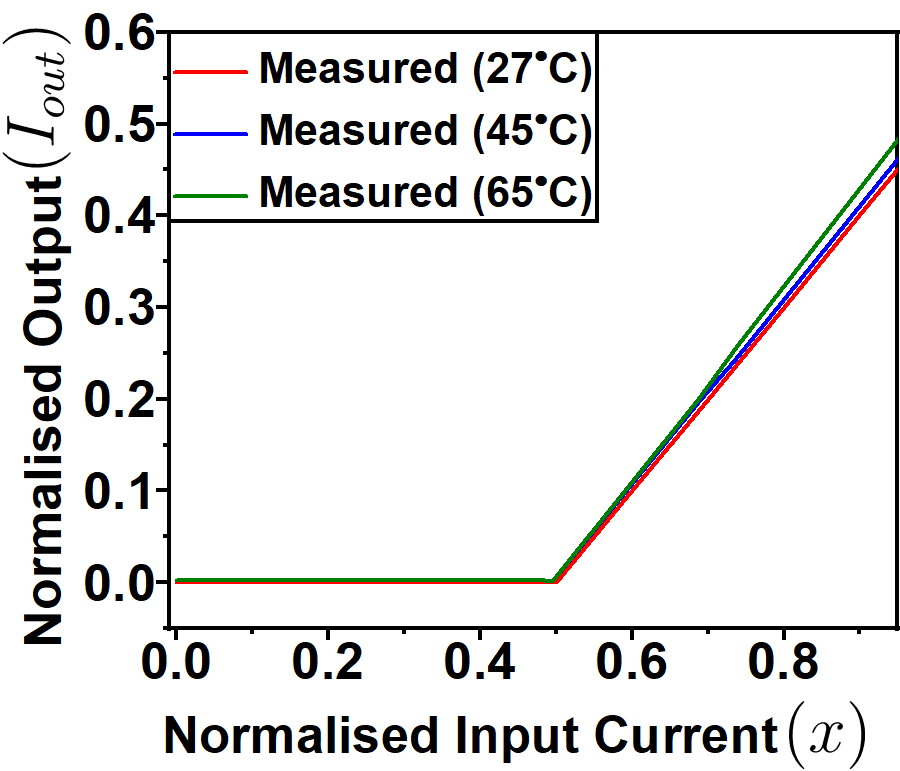}
	\label{temp1}}
	\subfloat[]{\includegraphics[width=0.317\linewidth]{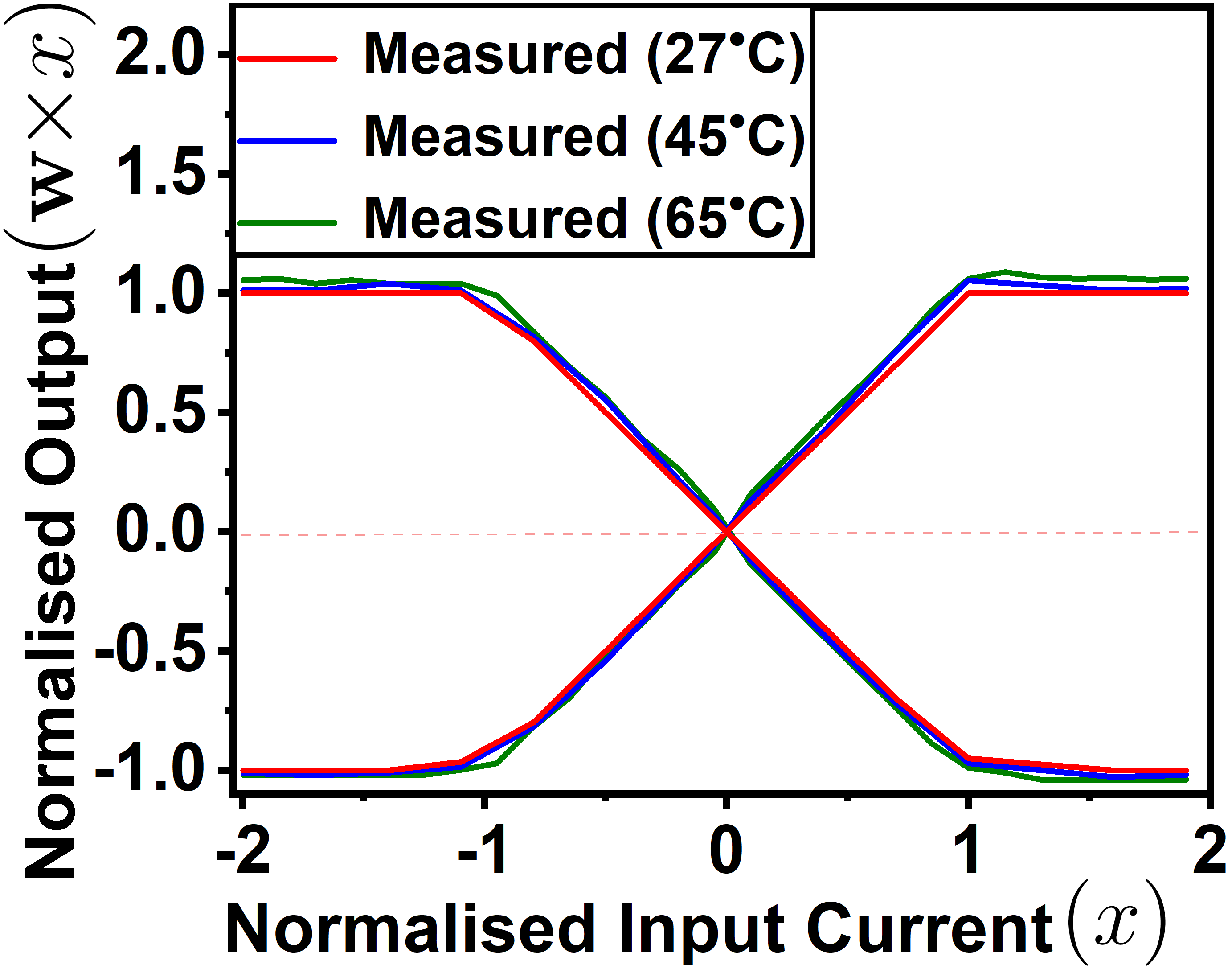}
	\label{temp2}}
	\subfloat[]{\includegraphics[width=0.31\linewidth]{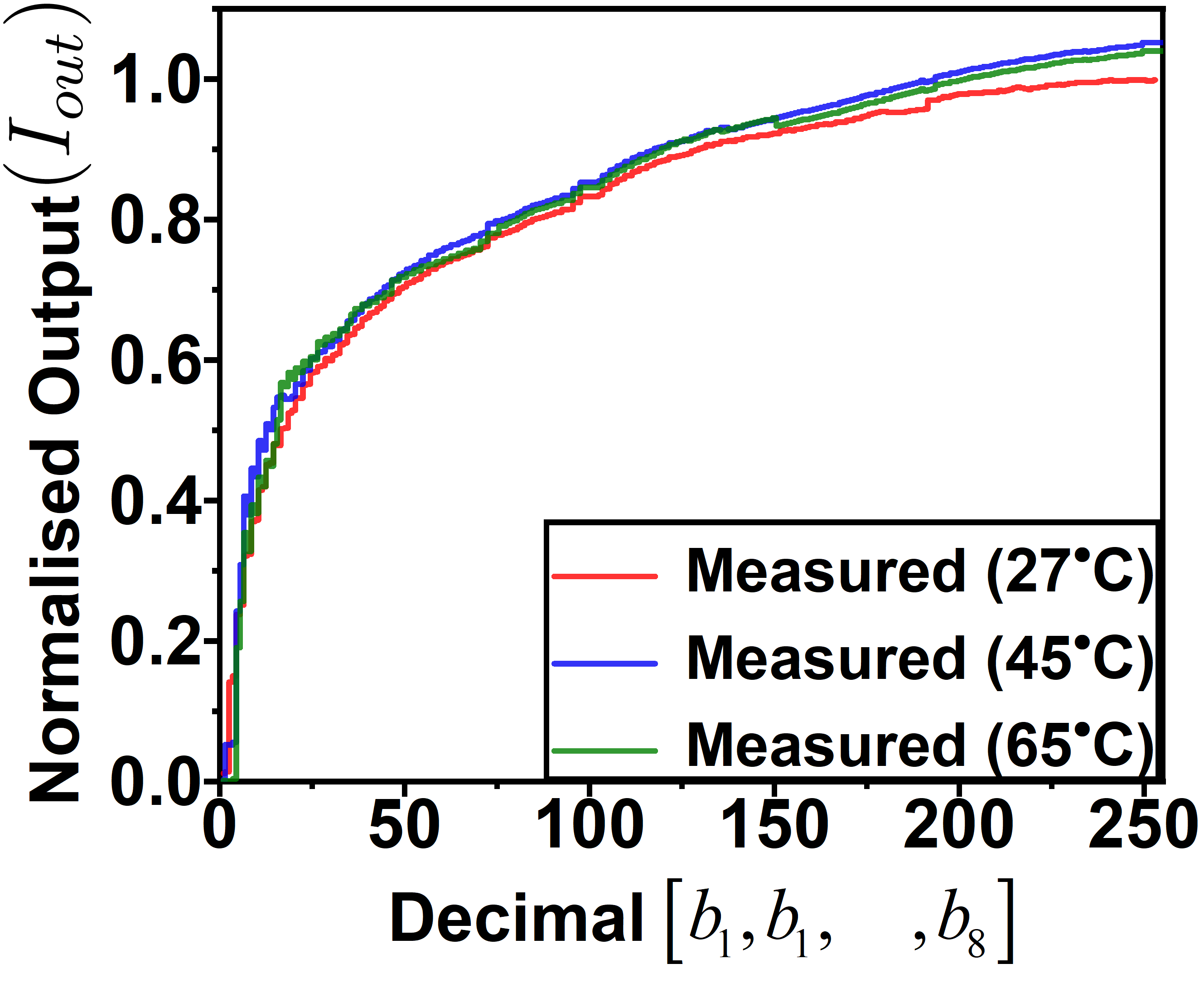}
	\label{temp3}}
 	\caption{ Measured charcterstics obtained across temperature for: \protect\subref{temp1} S-AC based soft ReLU; \protect\subref{temp2} S-AC based four-quadrant multiplier; \protect\subref{temp3} S-AC based 8-bit $log_2$ DAC.}
	\label{temp_figures}
\end{figure*}

\begin{figure*}[th]
\centering
\vspace{-7mm}
\subfloat[]{\includegraphics[width=0.32\linewidth]{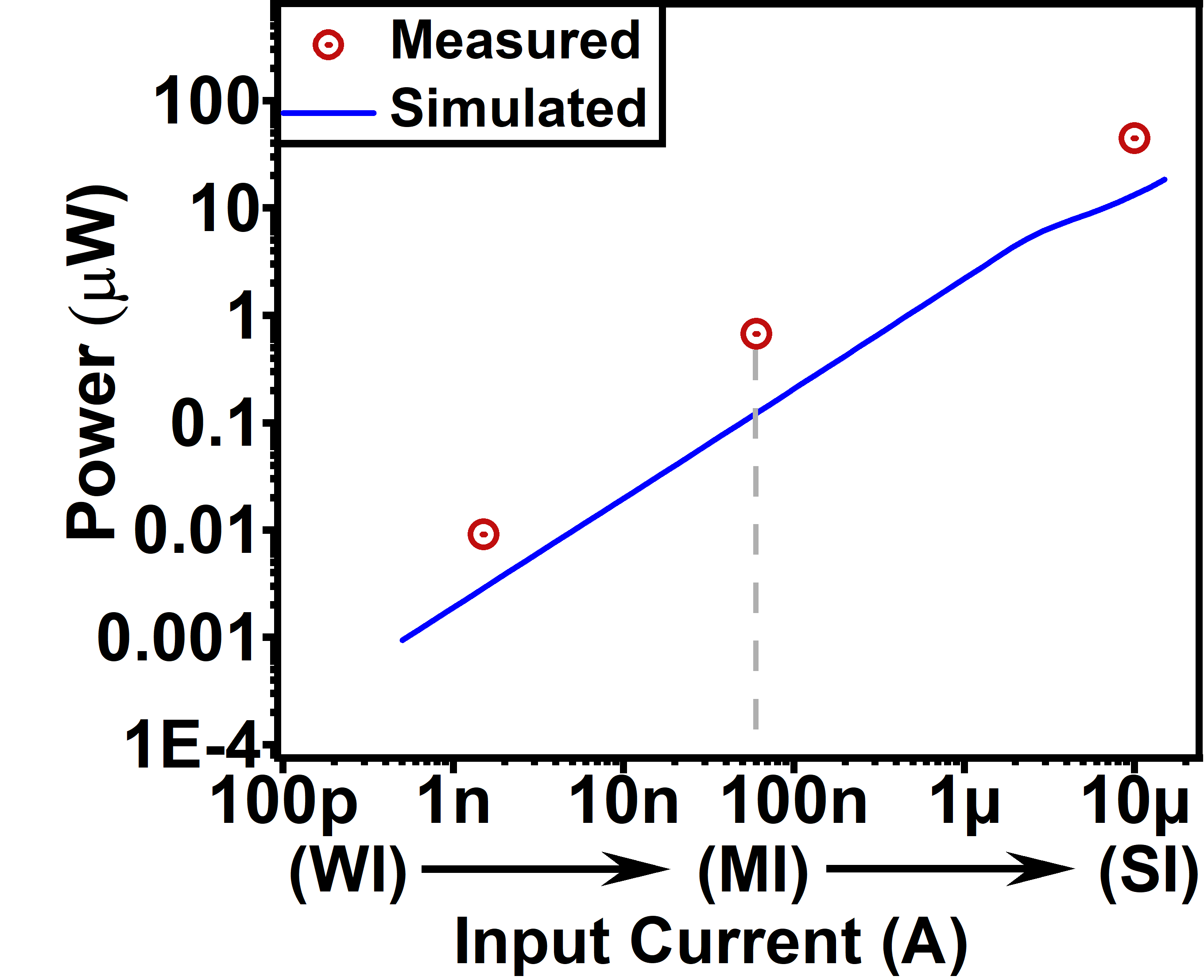}
\label{power_Sac}}
\subfloat[]{\includegraphics[width=0.317\linewidth]{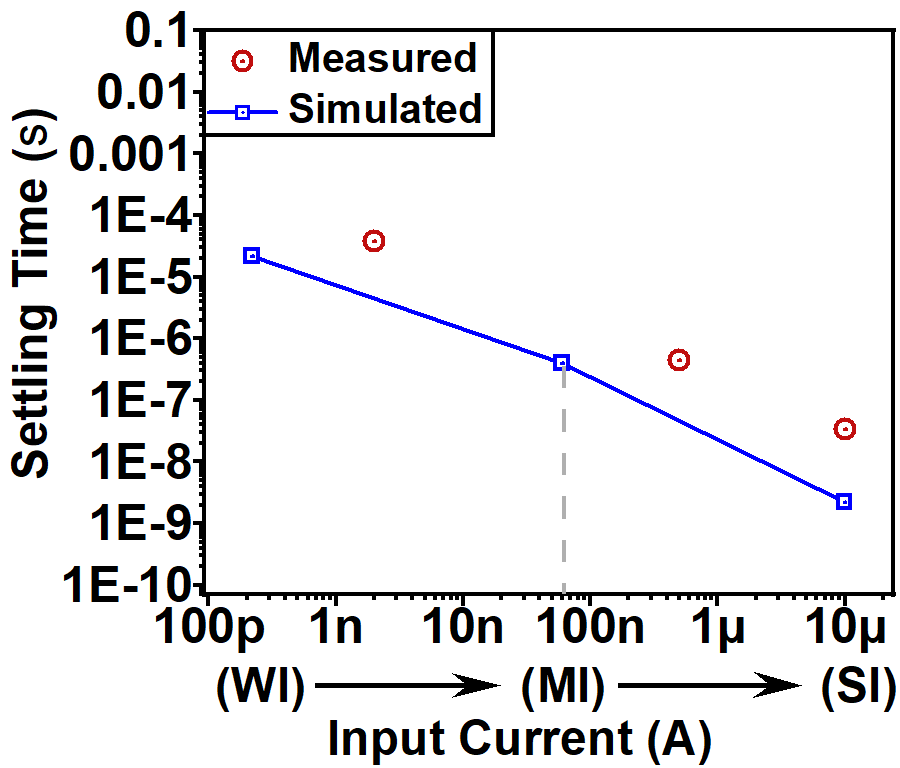}
\label{set_Sac}}
\subfloat[]{\includegraphics[width=0.34\linewidth]{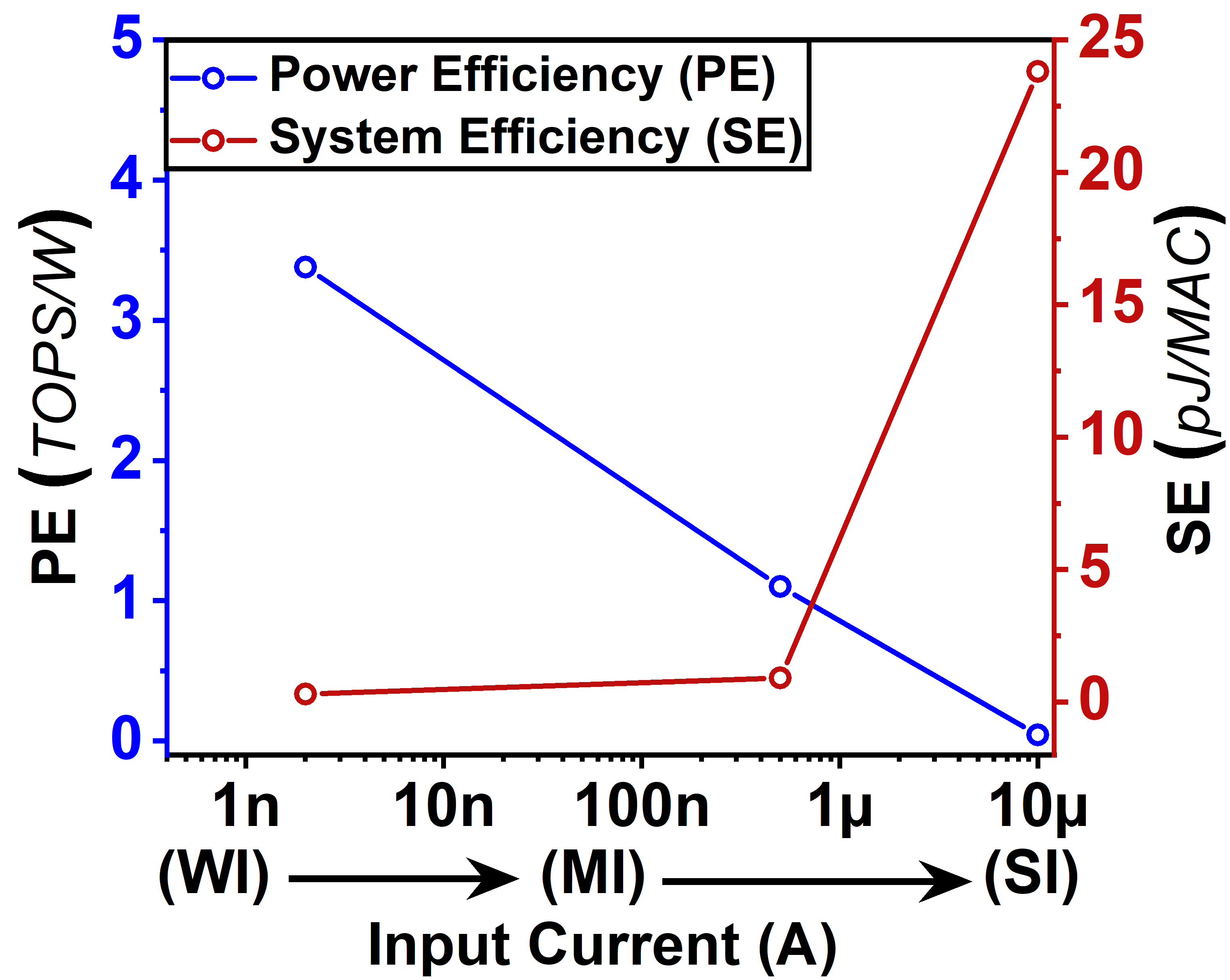}
\label{ce_se}}
\caption{Plots showing; \protect\subref{power_Sac}~average power consumption, \protect\subref{set_Sac}~settling time; \protect\subref{ce_se}~Performance efficiency and System efficiency of an S-AC unit biased in different operating regimes. }
\end{figure*}

\subsection{Design Margin and Shape Analysis} The shape-based analog computing framework is designed to preserve its transfer function within a stringent error margin under different biasing conditions (WI, MI, and SI)  and temperature variations. This, in turn, suggests that irrespective of mismatches between the drain-source currents or the gate-source voltages for a matched pair of MOS, the proto-shape ($h(\cdot)$ in Fig.~\ref{shape1_4spline}) remains intact. This proto-shape is considered important in machine learning applications and is governed by the design parameter $S$, which also decides the ability of the system to replicate the desired functional shape closely. S-AC design also relaxes the bounds of precise computing by allowing the user to choose the proto-shape (by choosing design parameter $S$) as per the need of the application and focus on obtaining desired functional shapes rather than conventional design techniques. Furthermore, like digital designs, the S-AC designs allow the user to trade off computational precision (by varying $S$) with energy and area~\cite{theory_sac}. On analysis, it was found that the design parameter $S=3$ is good enough to match most of the desired shapes with $>98\%$ accuracy, while even with the design parameter $S=1$ and $S=2$, the classification system accuracy does not drop significantly as the network learns with hardware approximation. 


\begin{table}[t]
\centering
\caption{Energy per Operational Unit}
\label{energy}
\resizebox{7.5cm}{!}{%
{
\begin{tabular}{|c|c|cc|}
\hline
\multirow{2}{*}{@   S=1, VDD= 1.1V} & \multirow{2}{*}{\begin{tabular}[c]{@{}c@{}}\# S-AC \\ Unit\end{tabular}} & \multicolumn{2}{c|}{\begin{tabular}[c]{@{}c@{}}Energy   Consumption\\  (pJ)\end{tabular}} \\ \cline{3-4} 
                                    &                                                                          & \multicolumn{1}{c|}{\hspace{1em} SI* \hspace{1.5em}}                              & WI*                              \\ \hline
S-AC                                & 1                                                                        & \multicolumn{1}{c|}{1.49}                             & 0.34                             \\ \hline
Multiply/Divide                     & 4                                                                        & \multicolumn{1}{c|}{5.23}                             & 1.19                             \\ \hline
Soft-ReLU                           & 2                                                                        & \multicolumn{1}{c|}{2.61}                             & 0.59                             \\ \hline
\end{tabular}}
}
 \begin{tablenotes}
 \item $^{*}$For operation in Strong Inversion (SI) regime Inversion Coefficient; $IC>10$ \& for Weak Inversion (WI) regime $IC<0.1$ \cite{binkley2007tradeoffs}.
 \end{tablenotes}
\end{table}

\subsection{Energy Analysis}
Table~\ref{energy} shows the best-case average energy consumed by S-AC based basic operational units. It can be noted that for strong inversion regime of operation, circuits were biased so as to  maintain the inversion coefficient, i.e., $IC>10$, while for weak inversion, $IC <0.1$ was maintained \cite{binkley2007tradeoffs}. This range of bias current corresponding to a particular operating regime was fed into the circuit to control the operating regime of the S-AC unit. It can be seen that as the circuit operating regimes move from SI to WI, the energy per operation decreases, while an optimal balance between power and speed will always be obtained in the MI regime.


\subsection{Performance analysis}
The most significant errors introduced in the operation of S-AC circuits are represented by mismatches, noise, and power-supply variations. As a result of these undesired effects, the functionality of the circuits can be severely affected by additive errors. In S-AC circuits, the margin between the shapes obtained in the SI and WI regimes takes into account all the variations due to second-order effects. This crucial feature allows the S-AC circuits to preserve the inherent shape of the implemented function.

\subsubsection*{Temperature variation} We compare the effect of nominal temperature variation on S-AC units. Fig.~\ref{temp_figures} shows the measured characteristic curves of S-AC based  ReLU, Multiplier, and DAC at different temperature points, respectively.  One can observe that even though there is a slight variation that can be attributed to the current mirrors in the desired curves, but the overall characteristic shape is preserved.

\subsubsection*{Power \& Task-Energy Efficiency} Fig.~\ref{power_Sac} shows a comparison plot between the measured and simulated power of S-AC based unit when the operating current is varied such that circuit operations move from WI to SI regime. It can be observed that the power consumption increases when circuit operation shifts from WI to SI regime.

\subsubsection*{Slew Rate} With the increase in the number of S-AC blocks, the corresponding slew rate and bandwidth increase as the number of inputs and the overall current available to charge the node capacitance increases. This results in an overall reduction in settling time and can be solely attributed to the constraints imposed by the hyper-parameter $C$ in~\eqref{char1_msmp}. It can also be noted that as the value of this hyper-parameter $C$ decreases, i.e., when the circuit operation shifts from SI to the WI regime, the settling time increases because it takes more time for the capacitor at the gate of the output transistor (node $V_B$ in Fig.~\ref{sbac_unit}) to charge with the limited available current.

\subsubsection*{Settling Time} This settling time (including dead time, slew time, and recovery time) decides the maximum input frequency at which the system can operate (assuming all the operations to be performed are done parallel) and can be given by (\ref{equation53})
\begin{align}
	{f_{max}} = \frac{1}{{\max \left( {{t_{settling,rise}},{t_{settling,fall}}} \right) + \Delta t}}
	\label{equation53}
\end{align}
Here, $\Delta t$ is the margin for the unexpected error that can arise due to circuit variations~\cite{pk2022hybrid}. It can safely be assumed to be between $5\%$ of $ t_{settling}$. Fig.~\ref{set_Sac} shows the measured settling time of an S-AC based unit when the operating current is varied such that the circuit moves from WI to the SI region of operation. It can be observed that as the operating regime moves from WI to SI, the time required to charge the capacitance node improves. Hence the circuit can operate at a higher speed. Fig.~\ref{ce_se} shows the variational performance efficiency $(PE)$ and system efficiency $(SE)$ when the circuit operating regime shifts from WI to SI. Note that $PE$ increases with the increase in operating current while $SE$ deteriorates as predicted in Fig.~\ref{tops}.

\begin{figure*}[th]
	\centering
	\subfloat[]{\includegraphics[width=0.9\linewidth]{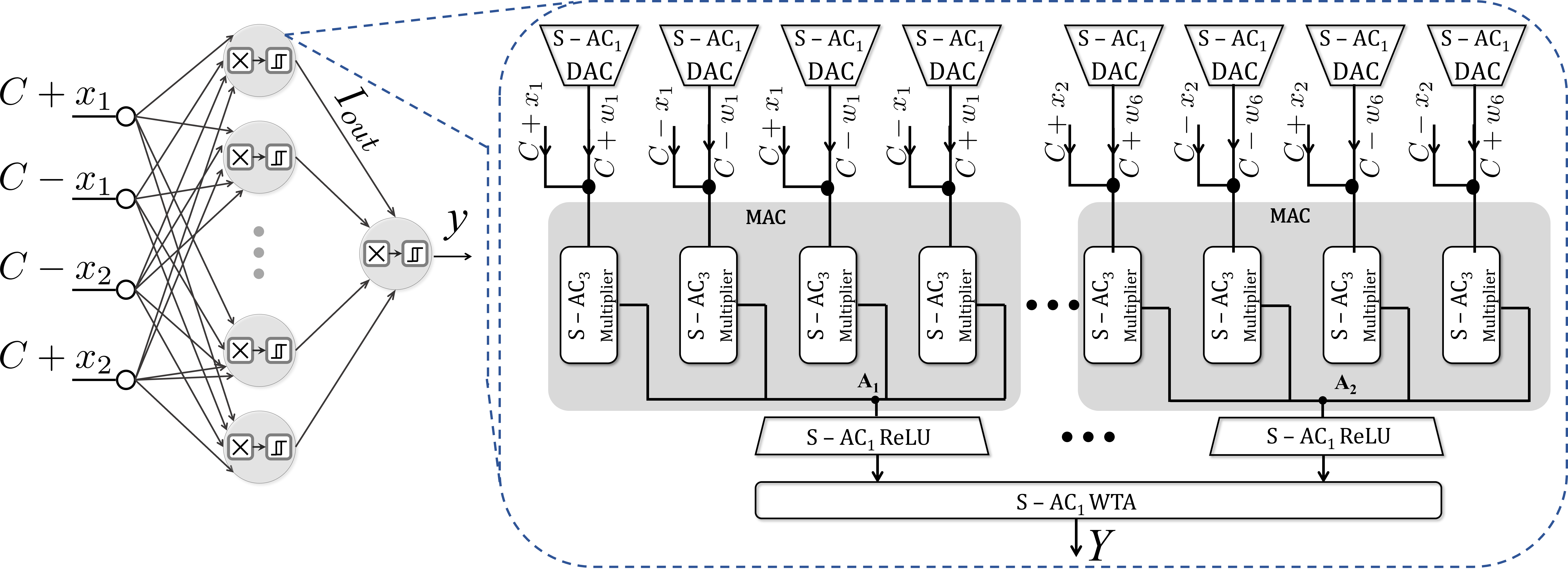}
	\label{application_architecture}}

\vspace{-4mm}
	\subfloat[]{\includegraphics[width=0.32\linewidth]{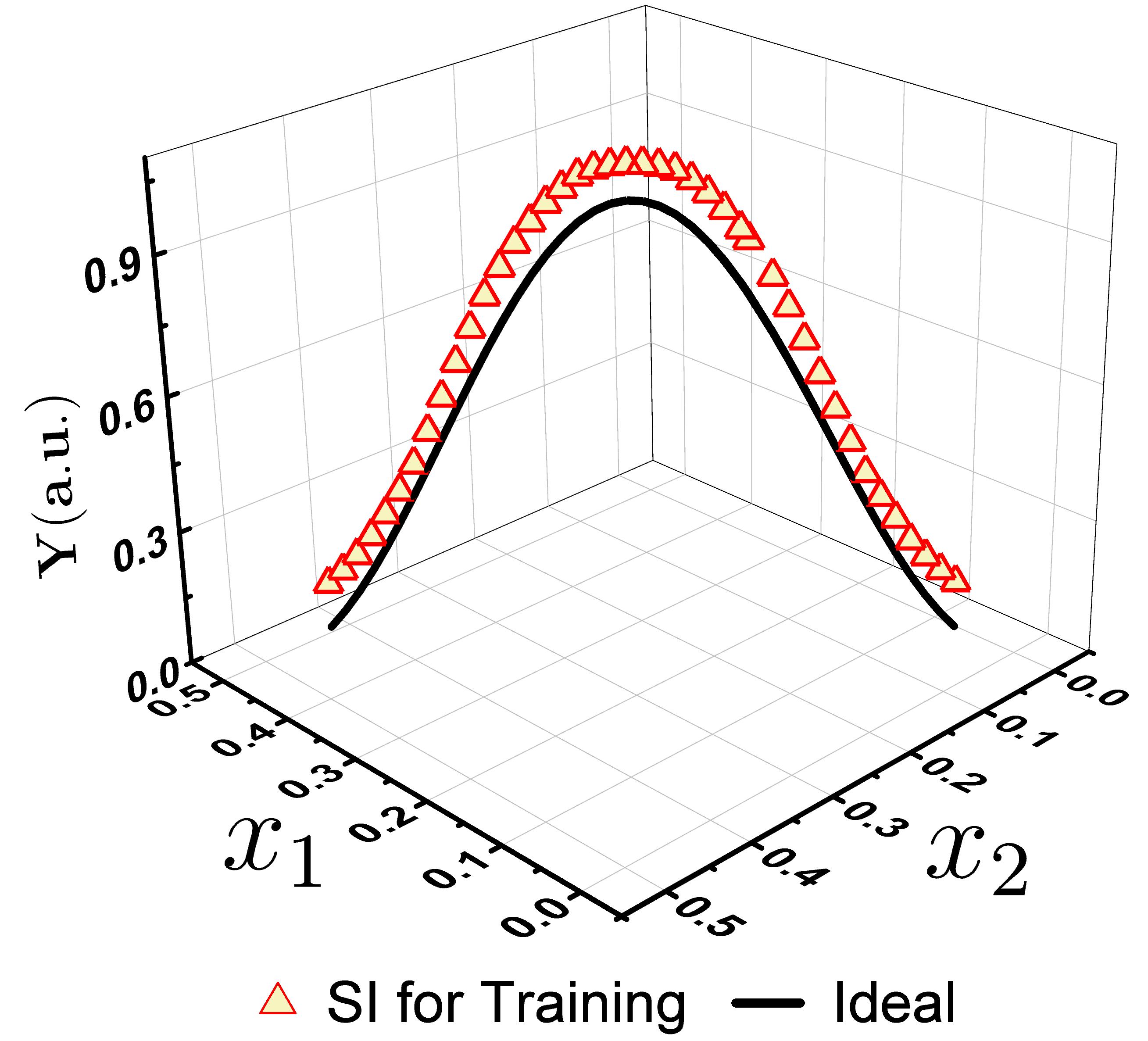}
	\label{sine_train}}
	\subfloat[]{\includegraphics[width=0.34\linewidth]{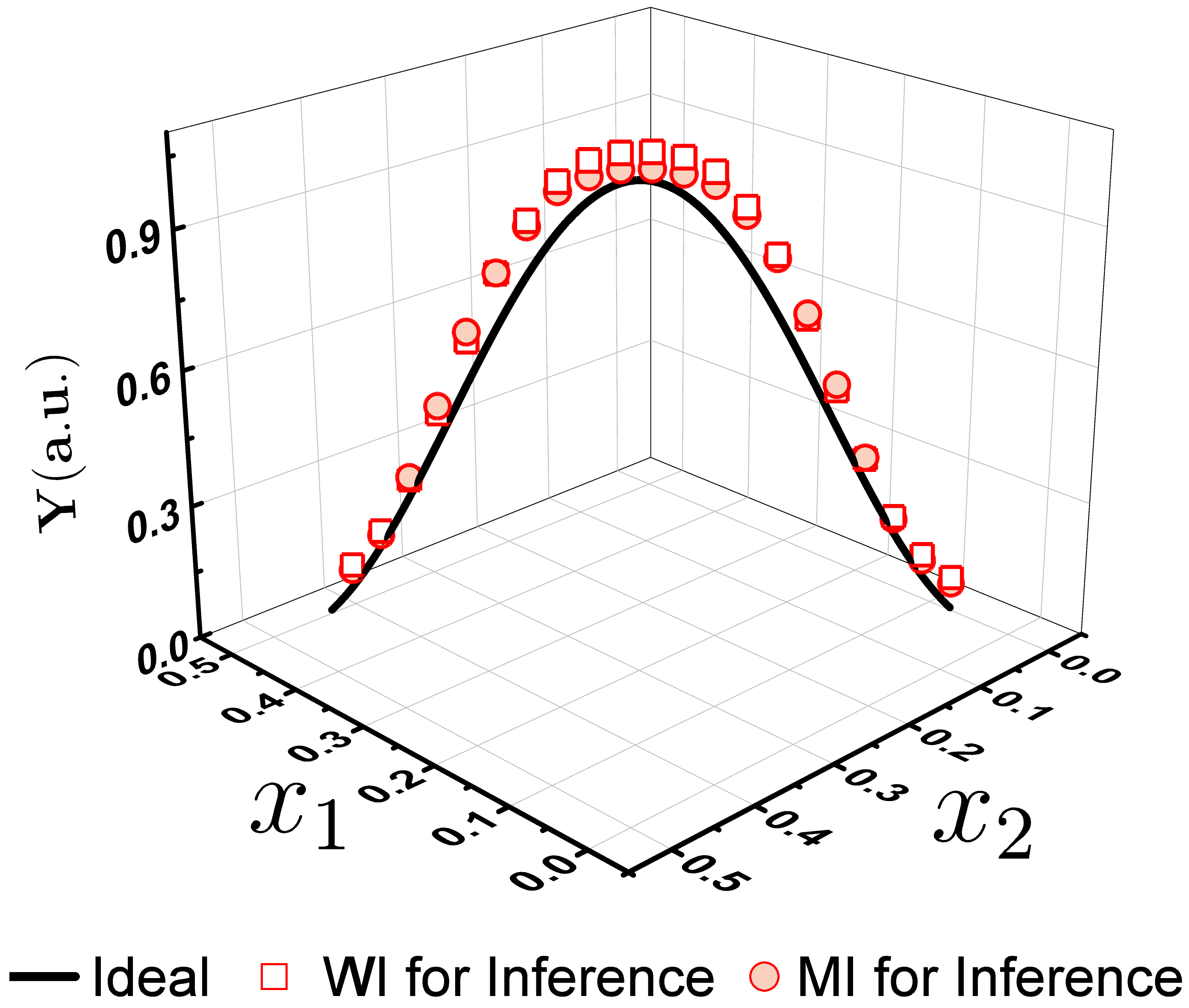}
	\label{sine_inf}}
	\subfloat[]{\includegraphics[width=0.34\linewidth]{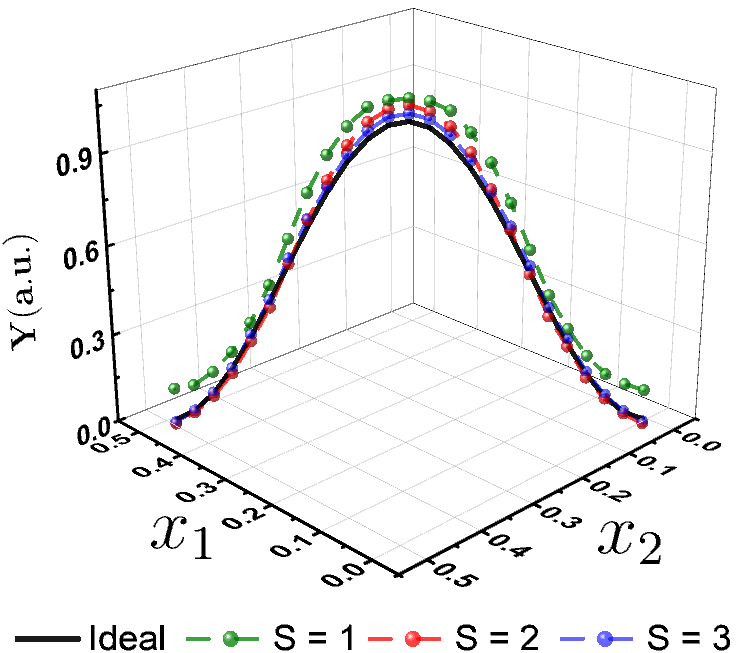}
	\label{multi_sine}}
 	\caption{\protect\subref{application_architecture}~System-level architecture of a 3-layer neural network, the inset shows the implementation of S-AC based near-memory core shown in Fig.~\ref{blockML_b} and implemented using S-AC DAC (Fig.~\ref{DAC_ckt}),  S-AC multiplier (Fig.~\ref{multiply_ckt}), S-AC ReLU (Fig.~\ref{nl_ckt}) and S-AC WTA \cite{theory_sac}; Measurement results of a Sine regression task performed on a 3-layer neural network showing: \protect\subref{sine_train}~training curve obtained in SI region and its comparison with ideal; \protect\subref{sine_inf}~inference curve obtained in MI and WI region and its comparison with ideal; \protect\subref{multi_sine}~Inference curve obtained for different design parameter $S$ in SI region and its comparison with ideal.}
	\label{app_figure}
\end{figure*}

\begin{table*}[t]
\centering
\captionsetup{labelformat=empty}
\caption{Comparison of S-AC based Analog Computational Blocks with Other Works}
\resizebox{\textwidth}{!} 
{
{
\begin{tabular}{|c|c|c|c|c|c|c|} 
\hline
\multicolumn{7}{|c|}{{\cellcolor[rgb]{0.937,0.937,0.937}}\textbf{Non-linearity}}                                                                                                                                                                                                                                                                                                                            \\ 
\hline
\textbf{Referred Work}                                                          & \cite{priyanka2018cmos}                                                                                   & \cite{zhu2019analog}                                                                                  & \cite{geng2020analog}             & \cite{krestinskaya2020memristive}                                                        &   \cite{synth_bias_scale}         &  $^{**}$This Work \\ 
\hline
\textbf{Implementation}                                                          & \begin{tabular}[c]{@{}c@{}}ReLU / Parametrised \\ ReLU\end{tabular} & ReLU                                                              & ReLU              & \begin{tabular}[c]{@{}c@{}}ReLU / Leaky \\ ReLU\end{tabular}        & ReLU      & Soft-ReLU                                                                \\ 
\hline
\textbf{Design based on}                                                         & Voltage mode                                                        & Voltage mode                                                      & Voltage mode      & Voltage mode                                                        & Current mode          & Current mode                                                             \\ 
\hline
\textbf{Building Block}                                                          & Half-Wave Rectifier                                                & Divider, Inverter  & \begin{tabular}[c]{@{}c@{}}Common Source\\ Amplifier\end{tabular}  & \begin{tabular}[c]{@{}c@{}}Transmission \\ Gate\end{tabular}                                                    & S-AC Unit         & S-AC Unit                                                                \\ 
\hline
\textbf{$^{*}$Operating Regimes}                                                       & SI                                                                  & SI                                                                & SI                & $-$                                                                 &  SI         & WI, MI, SI                                                               \\ 
\hline
\textbf{Technology ($nm$)}                                                    & 180                                                                & 55                                                               & 600             & 180                                                                & 7         & 180                                                                     \\ 
\hline
\textbf{Area ($\mu m^2$)}                                                        & $-$                                                                 & $-$                                                               & $-$               & 23.65 / 35.80                                                       & 1.06         & 190.46                                                                   \\ 
\hline
\textbf{Supply ($V$)}                                                              & $-$3 to $+$2                                                        & 1.2                                                               & 2.5               & 2.5                                                                 & 0.7         & 1.1 to 1.8                                                               \\ 
\hline
\textbf{Power}                                                                   & $-$                                                                 & 14.4$\mu$W                                                                &  $-$       & 0.11nW/2.15mW                                                     & 1.2$\mu$W          & 18.2nW - 89.4$\mu$W                                                      \\ 
\hline
\textbf{Result Type}                                                             & Simulated                                                           & Simulated                                                         & Simulated         & Simulated                                                           & Simulated & Measured                                                                 \\ 
\hline
\multicolumn{7}{|c|}{{\cellcolor[rgb]{0.937,0.937,0.937}}\textbf{Analog Multiplier}}                                                                                                                                                                                                                                                                                                            \\ 
\hline

\textbf{Referred Work}                                                          &\cite{ml3}                                                                                  & \cite{ml4}                                                                                  & \cite{al2015new}             & \cite{aloui2017cmos}                                                           &   \cite{synth_bias_scale}      & $^{**}$This work                                                     \\ 

\hline
\textbf{Design based on}                                                         & Current mode                                                        & Voltage mode                                                      & Current mode      & Current mode                                                        & Current mode          & Current mode                                                             \\ 
\hline
\textbf{$^{*}$Operating Regimes}                                                       & $-$                                                                 & WI                                                                & $-$                 & WI                                                                  & SI         & WI, MI, SI                                                               \\ 
\hline
\textbf{Technology ($nm$)}                                                    & 180                                                               & 180                                                              & 180              & 180                                                                & 7         & 180                                                                     \\ 
\hline
\textbf{Area ($\mu m^2$)}                                                        & 600/800                                                             & $-$                                                               & 147               & $-$                                                                 & 2.41         & 885.74                                                                   \\ 
\hline
\textbf{Supply (V)}                                                              & 1.2                                                                 & 1.3                                                               & 1.5               & $\pm$ 0.75                                                          & 0.7         & 1.1 to 1.8                                                               \\ 
\hline
\textbf{-3dB Bandwidth}                                                          & 79.6 MHz/59.7 MHz                                                   & 14 kHz                                                             & 230 MHz           & 300 MHz                                                             & 404.8 MHz         & 15.12 MHz                                                                \\ 
\hline
\textbf{Power}                                                                   & 60$\mu$W/75$\mu$W                                                   & 234$\mu$W                                                         & 700$\mu$W         & 0.15 mW                                                             & 36.43$\mu$W          & 546nW - 268.2$\mu$W                                                      \\ 
\hline
\textbf{Result Type}                                                             & Simulated                                                           & Simulated                                                         & Simulated         & Simulated                                                           & Simulated & Measured                                                                 \\ 
\hline
\multicolumn{7}{|c|}{{\cellcolor[rgb]{0.937,0.937,0.937}}\textbf{Logarithmic DAC}}                                                                                                                                                                                                                                                                                                                          \\ 
\hline
\textbf{Referred Work}                                                          & \cite{dac1}                                                                                   & \cite{dac2}    &  \cite{zhang2017memory}                                                                                & \cite{memristor_adc}            & \cite{dac4}                                                                & $^{***}$This Work                                                     \\ 
\hline

\begin{tabular}[c]{@{}c@{}}\textbf{Conversion }\\\textbf{Technique}\end{tabular} & \begin{tabular}[c]{@{}c@{}}Current\\ attenuator\end{tabular}        & \begin{tabular}[c]{@{}c@{}}Pseudo\\ -log amplifier\end{tabular}  & \begin{tabular}[c]{@{}c@{}}Weighted\\current source\end{tabular}   & Memristors        & \begin{tabular}[c]{@{}c@{}}Sub-threshold \\ transistor\end{tabular}         & \begin{tabular}[c]{@{}c@{}}S-AC based\\\end{tabular}                     \\ 
\hline
\textbf{$^{*}$Operating Regime}                                                        & $-$                                                                 & WI         & SI                                                        & $-$               & WI                                                                          & WI, MI, SI                                                               \\ 
\hline
\textbf{Technology ($\mu m$)}                                                    & 1.2                                                                 & 0.18             & 0.13                                                  & 0.18              & 0.18                                                                        & 0.18                                                                     \\ 
\hline
\textbf{Area ($mm^2$)}                                                           & 1.54                                                                & 1.5     & $-$                                                          & $-$               & 0.0069                                                                       & 0.00127                                                                  \\ 
\hline
\textbf{Supply (V)}                                                              & 5                                                                   & 1.65            & 1.2                                                  & 1.8               & 1.8                                                                          & 1.1 to 1.8                                                               \\ 
\hline
\textbf{Resolution (bit)}                                                        & 8                                                                   & 4      & 5                                                            & 4                 & 8                                                                           & 8                                                                        \\ 
\hline
\textbf{Usage}                                                                   & Log. DAC                                                            & Log. DAC                                                          & Log. DAC          & Log. DAC                                                            & Log. DAC          & \begin{tabular}[c]{@{}c@{}}Log. Compressive\\ DAC\end{tabular}           \\ 
\hline
\textbf{Power}                                                                   & 6mW @1MHZ                                                           & 13.2mW @1KHz    & $-$                                                  & 100$\mu$W @100kHz & 3.11$\mu$W @5MHz                                                             & \begin{tabular}[c]{@{}c@{}}138nW$-$ 536.4$\mu$W\\ @3.37MHz\end{tabular}  \\ 
\hline
\textbf{Result Type}                                                             & Measured                                                            & Measured                                                          & Measured         & Measured                                                            & Simulated & Measured                                                                 \\
\hline
\end{tabular}}
}
\label{comp_table}
 \begin{tablenotes}
 \item $^{*}$ For operation in Strong Inversion (SI) regime $IC>10$; for Moderate Inversion (MI) regime $0.1<IC<10$ and for Weak Inversion (WI) regime $IC<0.1$ was maintained~\cite{binkley2007tradeoffs}.
 \item $^{**}$ Measured results were performed for design parameter $S = 1$.
 \item $^{***}$ Log. Compressive DAC was designed for design parameter $S = 3$.
 \end{tablenotes}
\end{table*}

\section{Regression Results}\label{appl} 

In this section, we demonstrate the functionality of the S-AC based 3-layer neural network on a simple regression task. Fig.~\ref{application_architecture} shows the neural architecture of a S-AC based 3-layer neural network containing 6 hidden nodes and its corresponding circuit implementation. Here each node implements an analog ML core as shown in Fig.~\ref{blockML_b}. Here, S-AC compressive memory units are used to store weights in the compressed log domain. This near-memory computing architecture reduces the energy wasted in moving  data to and from the memory while simultaneously can operate at different bias currents. The parallel-connected S-AC multipliers whose outputs converge into a single node is representative of the proposed S-AC based Multiply-and-Accumulate (MAC) operation. However, the explanation and detailed implementation of S-AC based MAC and training methodology is beyond the scope of this work and shall be surfaced in the upcoming literature. The inputs to the architecture are first converted into differential compressive form and then passed to the hidden nodes. For demonstration, we use the S-AC architecture to learn a two-dimensional non-linear function given by 
\begin{align}
	Y = \sin \left( {2\pi {x_1}} \right) \sin \left( {2\pi {x_2}} \right)\
	\label{sine_equation}
\end{align}

The network was trained keeping in account the device mismatch obtained from post-layout simulation. Fig.~\ref{sine_train} shows the training curve obtained in the SI regime for $S=1$ and its comparison with the ideal curve. We further show in Fig.~\ref{sine_inf} that when the same architecture is used for testing while operating in the WI or MI regime, we are able to achieve nearly similar plots. Thus, we show how bias scalability can be used as an advantage to perform high-speed training in SI, whereas testing is done for low-power in WI using the same hardware. Fig.~\ref{multi_sine} shows the Sine regression curve obtained for varying design parameters $S$. We trained the network using the algorithm mentioned in \cite{multiplier_abhishek}. It can be observed that by increasing the design parameter $S$, a much closer approximation to the ideal curve is obtained. The mean square error (MSE) obtained between the ideal and desired curve decreases from $0.00781$ for $S=1$ to $0.00034$ for $S=3$. 

Table~\ref{comp_table} compares the measured performance of S-AC based analog computing blocks presented in this work with similar designs reported in the literature. It can be observed that the implemented designs can function over wide range of operating conditions (WI, MI, and SI) and at different power requirements with minimal area consumption when compared with similar technology node implementation. This is important to maintain the bias scalability of the designed circuit. This table shows the comparison of implemented ReLU activation with different variants of ReLU activations present in the literature. Table~\ref{comp_table} also shows the comparison of this work with various four-quadrant full precision and approximate analog multipliers present in the literature. In addition, a comparison of log compressive DAC present in this work is also done with other DAC implementations present in literature whose inherent characteristic response is logarithmic in nature. 

\section{Conclusion}\label{conc}

In this work, we proposed S-AC based bias-scalable analog computing processor and near-memory S-AC core for machine learning (ML) tasks. We reported the basic building blocks (S-AC compressive memory DAC, S-AC ReLU, and S-AC multiply-accumulate) of the S-AC core and also showed the implementation of S-AC based compressive memory DAC, which also mimics the computation using Bfloat16 and IEEE 754 single-precision number systems. As a proof of concept, we demonstrated the implementation of a 3-layer S-AC neural network performing standard ML regression at different biasing conditions (signifying different operating speed and power consumption) and for different design parameter $S$ (signifying different computational accuracy).

In addition, S-AC based analog computing blocks were shown to remain invariant to biasing conditions and and operating temperature. As a result, the S-AC based near-memory ML processor is well suited for high-speed training in the SI regime as well as for energy-efficient inference in WI regime, thereby allowing near-memory S-AC architectures to be used for both server and edge applications. It can be noted that a trade-off between speed and power can always be achieved by biasing in MI regime. At a system level, the overall performance (power and speed) of the S-AC processor can be adjusted by adjusting the hyper-parameter $C$ along with the input range, which in turn will bias the transistors in different operating regimes. In addition the design parameter $S$ allows the user to trade-off computational accuracy with the area and power~\cite{theory_sac}. We believe that this methodology can further be used to speed up other sub-variants of training algorithms such as physics-aware training \cite{wright2022deep} and hardware-algorithm co-design techniques \cite{lee2020hardware} to provide a boost in overall systems efficiency. Our future works will include the demonstration of generic programmable architecture for deep neural networks.

\begin{appendices}
\section{Proof: S-AC Implementation of Analog Multiplier}
\label{Appendix:A}
Consider the following equation, where $y$ is given by 
\begin{multline}
y = h(C + w + C + x) - h(C + w + C - x) +  \ldots \\
h(C - w + C - x) - h(C - w + C + x)
\label{exp0}
\end{multline}
The goal is to implement scalar multiplication between two variables $x$ and $w$. Here ${x,w,y} \in$ $\mathbb{R}$, $h$ is a non-linear monotonic function and $C$ is a hyperparameter. If we write the Taylor expansion of $h(x)$ around $w$ and ignore the higher-order terms, we will get
\begin{multline}
        h(C + w + C + x) = h(C + w) + (C + x)\Delta h(C + w) +  \ldots \\
    \frac{{C + {x^2}}}{2}{\Delta ^2}h(C + w)
    \label{exp1}
\end{multline}
\begin{multline}
        h(C+w+C-x) = h(C+w) + (C-x)\Delta h(C+w) + \ldots \\
    \frac{{{C-x}^2}}{2}{\Delta ^2}h(C+w)
    \label{exp2}
\end{multline}
\begin{multline}
        h(C-w+C-x) = h(C-w) + (C-x)\Delta h(C-w) + \ldots \\
    \frac{{{C-x}^2}}{2}{\Delta ^2}h(C-w)
    \label{exp3}
\end{multline}
\begin{multline}
        h(C-w+C+x) = h(C-w) + (C+x)\Delta h(C-w) + \ldots \\
    \frac{{{C+x}^2}}{2}{\Delta ^2}h(C-w)
    \label{exp4}
\end{multline}
Substituting \eqref{exp1} - \eqref{exp4} in \eqref{exp0} we get,
\begin{align}
y \cong \left[ {2x\Delta h(C + w) - 2x\Delta h(C - w)} \right]
\end{align}
\begin{align}
y \cong 2x\left[ {\frac{{dh(C + w)}}{{dw}} - \frac{{dh(C - w)}}{{dw}}} \right]
\end{align}
\begin{align}
y \cong 2x\left( {{w^ + } - {w^ - }} \right)
\end{align}
\begin{align}
    y \cong 2x \times w
    \label{multipliyeqn}
\end{align}
\end{appendices}

\section*{Acknowledgment}
The authors would like to acknowledge the joint IISc-WashU MoU to facilitate the collaboration between the two institutions. This work is also supported by the Department of Science and Technology of India (SERB CRG/2021/005478, DST/IMP/2018/000550). 

\bibliographystyle{IEEEtran}
\bibliography{ref}
\newpage

\vfill
\end{document}